# Magnetic energies (and some other parameters) in solar active regions of different Hale and McIntosh classes: statistical analysis for 2010-2024

I.V. Zimovets[1]*, I.N. Sharykin[1], W.-Q. Gan[2,3]

[1]Space Research Institute of the Russian Academy of Sciences, 84/32 Profsoyuznaya Str., Moscow, Russia, 117997

[2]Key Laboratory of Dark Matter and Space Astronomy, Purple Mountain Observatory, Chinese Academy of Sciences, Nanjing 210023, People's Republic of China

[3]University of Chinese Academy of Sciences, Nanjing 211135, People's Republic of China

*ivanzim@cosmos.ru

A statistical analysis of magnetic energies of the nonlinear force-free ($E_{NLFFF}$) and potential ($E_{POTF}$) fields, and their difference (a proxy for the free magnetic energy, $E_{FREE}$) in active regions (ARs) on the Sun of different Hale (Mount Wilson) and McIntosh classes for the period from 1.V.2010 to 12.VI.2024 is presented. The magnetic fields in ARs are calculated using the GX Simulator based on the information about ARs contained in the daily Solar Region Summary (SRS) files provided by the NOAA SWPC and vector magnetograms by the Helioseismic and Magnetic Imager (HMI) onboard the *Solar Dynamics Observatory* (SDO). Total unsigned and signed magnetic fluxes and vertical electric currents on the photosphere are also calculated. For the parameters considered, distributions have been determined in total for all ARs and separately for each Hale and McIntosh class; minimum, maximum, mean values of the parameters and standard deviations were calculated for each class. The magnetic energies, unsigned magnetic flux, unsigned vertical current, as well as the integral number of sunspots, number of ARs, and area of sunspots, integrated over ARs visible per day on the solar disk, exhibit similar ≈11.6-year cyclicity. On average, magnetic energies of ARs increase with increasing Hale and McIntosh class, while the average fraction of the free magnetic energy in ARs of different classes differs weakly. We also found that the Poisson Flare Probabilities (*PFPs*) correlate with the parameters, and the Pearson's correlation coefficient *ccp* is up to 0.89. The results reveal relationships between various parameters of ARs and may be used in developing prediction of space weather effects.

1. INTRODUCTION

Active regions (ARs) are places in the Sun's atmosphere where the brightest phenomena of solar activity occur, including such potentially geoeffective ones as flares and coronal mass ejections, CMEs [1, 2]. ARs are inextricably linked with groups of sunspots observed on the photosphere. In particular, the National Oceanic and Atmospheric Administration (NOAA) assigns a number to an AR when it contains at least one sunspot seen in the white-lite emission. We follow this approach to the definition of ARs in this paper. In the following, we will equate a sunspot group with an AR. However, it is necessary to keep in mind that a sunspot group is observed on the photosphere, while a corresponding AR is a limited three-dimensional space in various layers of the solar atmosphere, including the chromosphere, transition layer, and corona, above (and with a larger area than) this sunspot group. Magnetic field permeates and connects all these layers, and it is the main source of energy in ARs [3, 4]. Virtually all phenomena observed in ARs, such as hot coronal loops, plasma flows, jets, flares, CMEs, etc., are the result of the transformation of magnetic energy into the kinetic energy of plasma particles.

In many areas of science, it is useful to classify objects under study. This is valid for such objects as ARs on the Sun. Classifications allow describing the whole variety of ARs by a small number of observable (or extracted from observations) parameters. It facilitates cataloging and studying different physical processes and phenomena (e.g. solar flares) occurring in ARs. There are several different classifications of ARs. The two most famous and widely spread classifications of ARs (see [2]) are the Mount Wilson (or Hale, [5]) magnetic classification and the McIntosh morphological classification (e.g., [6]). The first classification contains four identifiers: α (a unipolar sunspot group), β (a simple bipolar sunspot group), γ (a group of sunspots with more complex distribution than of β-spots), and δ (a group of nearby sunspots with umbrae of opposite magnetic polarities situated within a common penumbra). The following classes are distinguished based on these identifiers: α, β, γ, βγ, δ, βδ, and βγδ, however the NOAA catalog for 2010-2024 (see below) does not contain sunspot groups with pure γ and δ classes, unlike some earlier works (e.g., [7, 8, 9]). The simplest magnetic class is α and the most complex class is βγδ. Further we will consider a simplified Hale classification with only five classes (α, β, βγ, βδ, and βγδ), which is currently widely spread (see examples of ARs of these five classes in fig. 1). The second classification divides sunspot groups into 60 classes of the type *Zpc*, where *Z* is the modified Zurich class, *p* is the type of the main sunspot, *c* characterizes the degree of compactness of the interior of a sunspot group [6]. Sunspot groups of the class *Axx* are the most compact and have the simplest magnetic configuration, while groups of the class *Fkc* are the largest and most complex. There are some other classifications of ARs, such as the magneto-morphological classification, MMC [10, 11], but in this work we restrict our consideration to the two classification systems of Hale and McIntosh only.

Works have been done showing the relationship between sunspot groups' classes and their flare activity. It has long been established that, on average, more complex and larger sunspot groups show higher flare productivity [12-15, 8]. For the Mount Wilson classification, the probability of flare occurrence grows from α through βγ to βγδ. For the McIntosh classification, the flare production rate increases along the diagonal line in the 3D parameter space from the simplest *Axx* class to the most complex *Fkc* class [16]. These results are confirmed by more recent statistical studies (e.g., [9, 17-22]). Flare productivity further increases as an AR evolves from a lower to higher McIntosh class over time [23, 24]. Moving from the classes to physical parameters of sunspot groups, a correlation was found between flare productivity of groups and area of their sunspots [12, 14, 23], as well as the magnetic flux [15, 8]. Rates of change of sunspot area [15] and magnetic flux [25, 26] in ARs also affect their flare productivity. However, the sunspot area and magnetic flux are indirect parameters for flare productivity. A more direct physical connection is expected between flare productivity and magnetic energy of ARs.

A great number of works are devoted to studying magnetic energy in ARs and its relations to flares and CMEs. Three types of magnetic energy in active regions are usually considered: 1) potential field energy, $E_{POTF}$, 2) linear or nonlinear force-free field energy, $E_{LFFF}$ or $E_{NLFFF}$, respectively, and 3) free magnetic energy, which is estimated as the difference between the first two energies, $E`_{FREE} \approx E_{LFFF} - E_{POTF}$ or $E_{FREE} \approx E_{NLFFF} - E_{POTF}$. It should be noted that magnetic field in the corona is not routinely measured, and in reality, magnetic field may differ from the force-free state [27]. Therefore, the magnetic energies considered should be treated with caution as approximate estimates. There are many detailed "case studies" of magnetic energies in individual ARs (e.g., [28-36]), and there are statistical studies that analyze magnetic energies in large samples (tens to hundreds) of ARs (e.g., [8, 37-43]). For example, it has been established that the dissipated free magnetic energy in ARs during powerful solar flares exceeds the kinetic energy of heated plasma, accelerated particles, CMEs and energy of emitted electromagnetic radiation in total. [38]. The energy dissipated during flares correlates well with the free magnetic energy, $E_{diss} \sim E^{0.9}_{FREE}$ [41]. Based on the analysis of several dozen events, it was found that the flare index moderately correlates (with a correlation coefficient from 0.55 to 0.76) with the free magnetic energy, and the correlation is comparable or a bit higher than correlation between the flare index and the magnetic flux [37, 40]. On the other hand, no significant correlation was found between the magnetic energies of an entire AR and the maximum class of a flare occurring in the AR within 24 h [43]. However, despite numerous studies and findings, to the best of our knowledge, no information on magnetic energy distributions in ARs of different Hale and McIntosh classes has been obtained so far.

The purpose of this work is to make a statistical study of magnetic energies $E_{POTF}$, $E_{NLFFF}$, $E_{FREE}$ (and some other parameters such as the total unsigned magnetic flux and vertical electric current) in ARs of various Hale and McIntosh classes for the period from 2010 to 2024 (≈1.3 solar cycles), in particular, to determine average values and standard deviations of these parameters in ARs of various classes. We will also estimate correlation coefficients between magnetic energies and flare productivity in ARs of different McIntosh classes, that has not been done previously. This study should add to the knowledge of ARs on the Sun and contribute to further understanding of why ARs of different classes exhibit different levels of flare productivity.

## 2. DATA AND METHODS

Information about ARs (sunspot groups) is taken from the daily Solar Region Summary (SRS) files [ftp://ftp.swpc.noaa.gov/pub/warehouse], namely: 1) NOAA AR number, 2) heliographic coordinates, 3) Carrington longitude ($Lo$), 4) sunspot group area ($A_{SS}$), 5) sunspot group longitude size ($LL$), 6) number of sunspots in a group ($N_{SS}$), 7) McIntosh class, and 8) Hale (Mount Wilson) class of a group.

The full-disk photospheric vector magnetograms of the *Helioseismic and Magnetic Imager* (HMI) on board the *Solar Dynamics Observatory* (SDO) [44, 45] with a time step of 12 min and 0.5" pixel resolution (0.91" optical resolution) were used to prepare the lower boundary conditions for extrapolation of magnetic field into the corona. We used only one set of magnetograms for each day obtained around 01.00 UT, which is close to the time of validity of SRS files (00.30 UT). In case of omissions in HMI data, the nearest set of magnetograms for a given day after 00.30 UT was taken. Additionally, we used nearby (in time) SDO/HMI optical light (pseudo-continuum around 6173 Å Fe I line) intensity maps with a step of 45 s for visual viewing and inspection of sunspot groups (see examples in fig. 1).

Extrapolation of magnetic field to the overlying layers of the solar atmosphere was made in approximations of the potential field (*POTF*) and the nonlinear force-free field (*NLFFF*) using the GX Simulator [46]. *POTF* is calculated from the normal magnetic component on the lower boundary (photosphere) with the fast Fourier transform method described in [47] and it is used as an initial condition for generating a *NLFFF* model with the optimization method proposed by [48, 49]. We used a rectangular Cartesian coordinate system with a spatial step of 1500 km (at *LL*≤*12°*) or 3000 km (at *LL*>*12°*) to reduce the calculation time. The size of the calculation area along the *X*-axis (longitude) was $L_x = k_x \times LL = L_{AR}$, along the *Y*-axis (latitude) $L_y = k_y \times LL$, and along the *Z*-axis (vertically above the photosphere) was equal to the larger of $L_x$ and $L_y$ (usually $L_x > L_y$). The coefficients $k_x$ and $k_y$ (with $k_x \geq k_y$) ranged from 2.4 to 12 depending on *LL*, the largest values were usually taken for ``unipolar'' ARs of α class to capture the distributed magnetic flux in a plage region around the principal sunspot (see fig. 1) (e.g., [1, 2, 50]). As a result of extrapolation for each AR for a given day, we obtained six three-dimensional data arrays: $B^x_{POTF}(x,y,z)$, $B^y_{POTF}(x,y,z)$, $B^z_{POTF}(x,y,z)$ and $B^x_{NLFFF}(x,y,z)$, $B^y_{NLFFF}(x,y,z)$, $B^z_{NLFFF}(x,y,z)$. From these six data arrays we calculated another three arrays: $B^x_{FREE}(x,y,z)$, $B^y_{FREE}(x,y,z)$, $B^z_{FREE}(x,y,z)$.

Magnetic energies of the potential and nonlinear force-free field were calculated numerically (in the CGS system of units) as a summation over voxels within the computational 3D domains

$$E_{POTF,NLFFF} \approx \sum_{i=0}^{i=N_i} \sum_{j=0}^{j=N_j} \sum_{k=0}^{k=N_k} \frac{B^2_{POTF,NLFFF}(i,j,k)}{8\pi} \Delta x \, \Delta y \, \Delta z \quad (1)$$

except 10% of voxels near the side and top boundaries of the three-dimensional domains to minimize boundary effects (see fig. 1). Here and below, indices *i, j, k* correspond to coordinates *x, y, z*, respectively. The proxy of the free magnetic energy in an AR was calculated as $E_{FREE}=E_{NLFFF}-E_{POTF}$. In addition to the magnetic energies, we also calculated for each AR the total unsigned magnetic flux

$$\begin{cases} TUFB_z \approx \sum_{i=0}^{i=N_i} \sum_{i=0}^{j=N_j} |B_z(i,j,k=0)|\Delta x \Delta y \\ |\boldsymbol{B}(i,j,k=0)| \geq B_{thr} = 220 \, G \end{cases} \quad (2)$$

and the total unsigned vertical electric current

$$\begin{cases} TUJ_z \approx \sum_{i=0}^{i=N_i} \sum_{i=0}^{j=N_j} |j_z(i,j,k=0)|\Delta x \Delta y \\ |j_z(i,j,k=0)| \geq 3\sigma(|j_z|) \\ |\boldsymbol{B}(i,j,k=0)| \geq B_{thr} = 220 \, G \end{cases} \quad (3)$$

on the photosphere. Here, the summation for the unsigned magnetic flux is performed using pixels on the photosphere in which the absolute value of the magnetic field exceeds the threshold value $B_{thr}=220$ G, which was used in [51] for HMI data. When summing up the unsigned vertical current, in addition, only pixels in which the absolute value of the vertical current exceeded the triple standard deviation (3σ) for the considered calculation region were taken into account. The vertical component of electric current density on the photosphere in an AR was standardly calculated using the theorem of magnetic field induction (Ampere's law) in the differential form (e.g., [52, 36])

$$j_z(i,j,k=0) \approx \frac{c}{4\pi}\left(\frac{\Delta B_x(i,j,k=0)}{\Delta y(i,j,k=0)} - \frac{\Delta B_y(i,j,k=0)}{\Delta x(i,j,k=0)}\right) \quad (4)$$

The calculations were made (with a similar procedure as in [52]) for the time interval from May 1, 2010 (the beginning of regular HMI observations) to June 12, 2024, for ARs in the central part of the solar disk. The longitudes and latitudes of the boundaries of the calculated ARs were limited within ±70° relative to the disk center. A total of 5144 daily SRS files were used. The NOAA AR numbers for the considered interval varied from 11064 to 13711, i.e., there were 2647 unique ARs. An AR with the same unique NOAA number could be present in the central part of the solar disk for several days, and Hale and McIntosh classes of such AR could change in time. For each day, we considered such an AR independently. For example, if an AR was observed for N days, we made N independent extrapolations and calculations for it and included it in our database N times. For this reason, we obtained 13414 sets of data arrays in total instead of 2647. Then we selected only those ARs (5301 out of 13414, or ≈40%), which met the following additional criteria: 1) the imbalance of positive and negative magnetic fluxes on the photosphere is no more than 20%, 2) the imbalance of positive and negative vertical electric currents on the photosphere is no more than 20%, and 3) the free magnetic energy in the AR takes a positive value, $E_{FREE}>0$. Files with numerical information about 5301 ARs selected for analysis and a brief description of the data structure are available at the following link https://disk.yandex.ru/d/OxvqZDE8-48r1A. The histograms of numbers of selected ARs of different Hale classes are shown in fig. 2a, and of different McIntosh classes - in fig. 2b.

We estimated errors of the calculated parameters ($E_{POTF}$, $E_{NLFFF}$, $E_{FREE}$, $TUFB_z$, $TUJ_z$), using the following Monte Carlo approach, which was applied, for example, in [53, 34]. For each AR shown in fig. 1 we created two-dimensional arrays of noise components $B^x_{noise}(x,y,z=0)$, $B^y_{noise}(x,y,z=0)$, $B^z_{noise}(x,y,z=0)$ using a random number generator and added them to the three corresponding components of the magnetic field vector on the photosphere at each pixel. The three noise components of the magnetograms had a normal distribution around zero with a standard deviation of 50 G. This value is comparable to the threshold of the transverse magnetic field component in full-disk HMI data [54], and it exceeds the nominal noise level of about 5-10 G for the line-of-sight component [51]. Extrapolations were made (with GX Simulator) using these "artificial" magnetograms and different parameters for ARs were calculated as described above. For each AR, this procedure was repeated 50 times, after which the mean values of the parameters and standard deviations were calculated. A relative error is defined as the ratio of the standard deviation to the mean value. The relative errors for $E_{POTF}$ ranged from 0.3 to 0.8%, for $E_{NLFFF}$ from 2.8 to 5.1%, for $E_{FREE}$ from 9.6 to 20.9%, for $TUFB_z$ from 0.2 to 0.5%, and for $TUJ_z$ from 1.7 to 8.6%. It was not feasible to apply this procedure to all 13414 ARs due to the cost of machine time (we estimated it would take more than 7 years for our personal computer). As a result, we chose error estimates of the parameters under consideration for the whole sample of ARs as the integer rounding of the highest values obtained, i.e. 1% for $E_{POTF}$, 5% for $E_{NLFFF}$, 21% for $E_{FREE}$, 1% for $TUFB_z$, and 9% for $TUJ_z$. Expectedly the free magnetic energy has the largest error. These error estimates should be treated with caution because they do not take into account many factors, for example, incorrect elimination of the π-ambiguity of the azimuthal magnetic field component, finiteness of the spatial integration steps, sphericity of the Sun's surface, and, most importantly, possible deviation of the real magnetic field from the force-free and potential state. The discussion of this complicated issue is beyond the scope of this paper.

## 3. RESULTS

### 3.1. EVOLUTION OF PARAMETERS INTEGRATED OVER ARs OF ALL CLASSES

Let us first consider the dynamics of the ARs' parameters (without division into classes) for the whole time interval under consideration from 2010 to 2024. The temporal profiles of the total sunspot number, number of ARs, sunspot area, magnetic energies, unsigned magnetic flux, and unsigned vertical electric current integrated over all ARs (within ±70° from the disk center) in 1-day steps are shown in fig. 3. As expected, all parameters are subject to variations of solar activity cyclicity, with a period of about 11.6 years. The parameters are correlated with the sunspot number, with the linear Pearson correlation coefficients (*ccp*) ranging from 0.65 to 0.88, which is shown in the bottom right corners on the corresponding panels. The *ccp* between the sunspot number and $E_{POTF}$, $E_{NLFFF}$,

$E_{FREE}$ are 0.77, 0.80, and 0.65, respectively. It is interesting to note that *ccp* between the sunspot number and the total unsigned vertical current (0.88) is even higher than between the sunspot number and the total sunspot area (0.81), and also it is higher than *ccp* between the sunspot number and the total unsigned magnetic flux (0.83).

### 3.2. STATISTICS FOR ARs OF HALE CLASSES

In terms of percentage, the ARs selected were distributed as follows according to the Hale magnetic classes (see Table 1 and fig. 2a): α – 9.6% (15.8%), β – 66.6% (59.6%), βγ – 15.9% (16.4%), βδ – 1.8% (2.6%), and βγδ – 6.1% (5.5%). The percentages of "unique" ARs are given in parentheses (i.e. when each NOAA AR is counted only once). Except for ARs of the class α, these values are similar to the values obtained in [55] for the interval 1992-2015. In our sample, the percentage of α ARs is lower than in [55]. This is likely because some "unipolar" α ARs were discarded by our algorithm since they did not meet the three additional criteria we used (see Section 2).

The distributions of $E_{POTF}$, $E_{NLFFF}$, $E_{FREE}$, the fractions $E_{FREE}/E_{NLFFF}$ and $E_{FREE}/E_{POTF}$, and their paired dependencies on each other for ARs of five Hale classes are shown in fig. 4. One can see:

1. on average, magnetic energies of ARs increase with increasing Hale class, from α to βγδ (fig. 4a-f),
2. while the average fraction of $E_{FREE}$ to $E_{NLFFF}$ or to $E_{POTF}$ in ARs of different Hale classes differs weakly (fig. 4g,h),
3. the free magnetic energy $E_{FREE}$ in ARs of the βδ class has two maxima, the left one is close to the maximum of the distributions for the lower classes α, β, βγ and the right one is close to the maximum for the highest class βγδ (fig. 4f), i.e. this class may be considered as intermediate in terms of $E_{FREE}$,
4. there is very high correlation between $E_{NLFFF}$ and $E_{POTF}$, the Pearson linear correlation coefficient *ccp* is almost identical for ARs of different Hale classes (from 0.96 to 0.99; fig. 4a),
5. $E_{FREE}$ correlates with $E_{POTF}$ and $E_{NLFFF}$ more weakly (*ccp* is from 0.32 to 0.71; fig. 4c,e).

Figure 5 shows the paired dependences of the magnetic energies $E_{POTF}$ (panel a), $E_{NLFFF}$ (panel b), $E_{FREE}$ (panel c) on five other important physical parameters for ARs of different Hale classes: the number of sunspots in an AR, $N_{SS}$ (line 1 from the top), the sunspot area, $A_{SS}$ (line 2), the longitude length of the AR in degrees, $L_{AR}$ (line 3), the total unsigned magnetic flux, $TUFB_Z$ (line 4), and the total unsigned vertical electric current, $TUJ_Z$ (line 5). There are correlations between the magnetic energies and the considered ARs' parameters, with stronger correlations with the parameters for $E_{POTF}$ and $E_{NLFFF}$, and weaker correlations for $E_{FREE}$ (that is probably because of the lower accuracy of $E_{FREE}$ determination). The strongest correlations (0.91 to 0.95) are between $E_{POTF}$, $E_{NLFFF}$ and $TUFB_Z$ (fig. 5a4,b4). There is no significant difference between the correlation coefficients for ARs of different Hale classes. The next highest correlations (0.70 to 0.88) are between $E_{POTF}$, $E_{NLFFF}$ and $A_{SS}$ (fig. 5a2,b2), and comparable correlations (0.63 to 0.84) are between $E_{POTF}$, $E_{NLFFF}$ and $TJB_Z$ (fig. 5a5,b5).

Relying on the obtained dependences (in fig. 5), we fit pairs of parameters for the full sample of ARs (without division into Hale classes) using the simple power-law model

$$y = 10^a x^b \qquad (5)$$

where $x$ is a given parameter ($N_{SS}$, $A_{SS}$, $L_{AR}$, $TUFB_Z$ or $TUJ_Z$) and $y$ is one of the magnetic energies ($E_{POTF}$, $E_{NLFFF}$ or $E_{FREE}$). The power-law models were used, e.g., in [41, 42, 43] to relate some AR' parameters with others. We determined the values of the model coefficients ($a$ and $b$) and estimated their errors by minimizing the chi-square error statistic with the least squares method. For this purpose, we used the parameters' errors described in Section 2. The fitting was done using the "Linfit.pro" function in IDL in the log-log parameter space. The obtained values of the models' coefficients, $a$ and $b$, and their errors (one standard deviation) are presented in Table 2. These power-law models can be used further (in other works) to estimate the magnetic energies of ARs based on the considered parameters, which are easier to obtain from observations than to extrapolate magnetic field and calculate the magnetic energies. For example, these models may help to make a retrospective analysis of the magnetic energies of ARs from optical observations of sunspots (without magnetic field measurements), although the accuracy will be lower than if one uses magnetic flux calculations from photospheric magnetograms because the correlation of magnetic energies with the magnetic flux is higher (see fig. 5).

The minimal, maximal, mean values, and standard deviations of all considered parameters (not only those ones discussed in the previous paragraph) for ARs of different Hale classes and all ARs without class separation are summarized in Table 1.

### 3.3. STATISTICS FOR ARs of MCINTOSH CLASSES

Let's move on to considering the parameters for ARs of different McIntosh classes. First, we plotted the graphs (fig. 6) of paired dependences of the magnetic energies ($E_{POTF}$, $E_{NLFFF}$, $E_{FREE}$) on each other and also the fraction of the free magnetic energy on the energy of the nonlinear force-free field for all considered ARs of different McIntosh classes (except 3 classes *Esc*, *Fro*, *Fri*, for which there were no ARs in our dataset), similar to how it was done for ARs of different Hale classes (see fig. 4a,c,e,g). Here, each AR is shown as a separate point, and the color of the point indicates its belonging to a certain McIntosh class, the designations of which are given to the right of the graphs. One can see from these graphs that the magnetic energies, in general, increase with the growth of the McIntosh class, from the simplest and compact *Axx* to the largest and most complex ARs of the *Fkc* class, while the fraction of the free magnetic energy does not show such a trend (similar to ARs of Hale classes).

The dependence of the magnetic energies and other AR's parameters on the McIntosh classes can be seen more clearly in fig. 7, which shows the average values of the parameters and their standard deviations (±1σ) for each class. The sunspot number increases relatively smoothly with increasing McIntosh class, showing small fluctuations, with the gap for "unipolar" classes *Hrx-Hkx*. Other parameters show a similar trend, but there are local peaks at a few intermediate classes, such as *Hhx-Hkx*, *Cho-Cki*, *Dhi-Dki*, *Eki-Ekc,* and *Fki-Fkc*. Probably these peaks are associated with the presence of a large principal sunspot in such ARs (middle letter *h* or *k* in the class name). It should be noted that the maximum average values of magnetic energies $E_{POTF}$, $E_{NLFFF}$, and $E_{FREE}$ are found in ARs of the top *Fkc* class among all McIntosh classes.

The minimal, maximal, mean values, and standard deviations of all considered parameters for ARs of different McIntosh classes are summarized in Table 3.

Finally, having average values of the parameters for the ARs of different McIntosh classes, let us consider their correlations with flare productivity. For this, we consider a convenient measure of flare productivity called the Poisson Flare Probability (*PFP*), the values of which for all the main McIntosh classes were obtained statistically in [18]. The *PFP* provides an estimate of probability that an AR of a given McIntosh class will produce at least one flare of a given X-ray class (in particular, C, M, or X) per day (24 h). Paired dependences of the *PFP* and average values (±1σ standard deviation) of eight ARs' parameters, including magnetic energies $E_{POTF}$, $E_{NLFFF}$, and $E_{FREE}$, are shown in Fig. 8. The left column is for class C flares, the middle column is for class M, and the right column is for class X flares. The color of the data points corresponds to the McIntosh class, from black for *Axx* to red for *Fkc* (as in Fig. 6). The Pearson linear correlation coefficient values (*ccp*) are shown in the upper left corner of the corresponding panels. The following can be noted. First, on average, the *PFP* values are higher for C and M class flares (left and middle panels, respectively) than for X class flares (the highest *PFP* value of X class flares is only 0.27 for the *Fkc* class). This can be explained by the higher frequency of C and M class flares and, accordingly, the higher accuracy of determining their *PFP*. Second, there are moderate correlations between the *PFP* and some of the parameters under consideration, the *PFP* is mostly higher than 0.5-0.6. However, for some parameters (and especially for M class flares), the correlation coefficient takes on large values up to 0.89 for the sunspot number, 0.84 for the total unsigned vertical electric current, and 0.83 for the free magnetic energy. The last two parameters are related to AR's "nonpotentiality" and free magnetic energy. The presence of a high correlation between the *PFP* and these parameters confirms their important role in flaring productivity of ARs. In other words, a larger and more complex AR contains, on average, stronger electric currents and higher free magnetic energy, and is more likely to produce powerful solar flares.

## 4. CONCLUSIONS

We performed the statistical study of three types of magnetic energies ($E_{NLFFF}$, $E_{POTF}$, and $E_{FREE}$), the fractions $E_{FREE}/E_{NLFFF}$, $E_{FREE}/E_{POTF}$ and several other parameters (sunspot number - $N_{SS}$, sunspot area - $A_{SS}$, longitudinal length - $L_{AR}$, total unsigned magnetic flux - $TUFB_Z$, total unsigned vertical electric current - $TUJ_Z$) in ARs observed in the central part of the solar disk in the period from 1.V.2010 to 12.VI.2024. A total of 13414 sets (each corresponds to one "isolated" AR) of three-dimensional arrays were obtained with the magnetic field extrapolated in the potential and nonlinear force-free approximations using the GX Simulator based on the SDO/HMI photospheric magnetograms and daily SRS (NOAA) files. From this set, 5301 subsets (about 40%) were selected that met several criteria. We determined minimum, maximum, mean values of the parameters and the standard deviations for all ARs in total and separately for ARs of each Hale (Mount Wilson) and McIntosh classes. These parameters are summarized in two tables (1, 3). As far as we know, this is the first time this has been done.

We found that magnetic energies, unsigned magnetic flux, unsigned vertical current, as well as the integral number of sunspots, number of ARs, and area of sunspots, integrated over ARs visible per day in the central part of the solar disk, exhibit similar ≈11.6-year cyclicity. On average, magnetic energies of ARs increase with increasing Hale class (from α to βγδ) and McIntosh class (from *Axx* to *Fkc*), while the average fraction of the free magnetic energy, $E_{FREE}/E_{NLFFF}$ (or $E_{FREE}/E_{POTF}$) in ARs of different classes differs weakly. For a sample of all ARs (without dividing into classes), using the least squares approximation, we determined the models of the power-law dependence of the magnetic energies $E_{NLFFF}$, $E_{POTF}$, and $E_{FREE}$ on the ARs' parameters $N_{SS}$, $A_{SS}$, $L_{AR}$, $TUFB_Z$, $TUJ_Z$ . The coefficients of these models with errors are summarized in a separate table (2) for convenience. We also found that the Poisson Flare Probabilities (*PFPs*) for ARs of different McIntosh classes correlate to some level with the ARs' parameters, and the Pearson's correlation coefficient *ccp* is up to 0.89. The results reveal statistical relationships between various parameters of solar ARs and may be used in developing prediction of severe space weather effects.

**ACKNOWLEDGMENTS**

The authors are grateful to the SDO/HMI (NASA) instrument team and SWPC NOAA for free access to the data, without which this work could not be carried out. SDO is the mission of NASA's Living With a Star (LWS) Program. We also thank the developers of the GX Simulator software package.

**FUNDING**

The work of I.V. Zimovets and I.N. Sharykin (preparation of data, extrapolation of magnetic field, calculation of magnetic parameters and its statistical analysis) was supported by a grant from the Russian Science Foundation (project no. 23-72-30002). The work of W.-Q. Gan was supported by the National Natural Science Foundation of China (12233012, 12333010, 11921003), the Strategic Priority Research Program of the Chinese Academy of Sciences (XDB0560000), and the National Key R&D Program of China (2022YFF0503002).

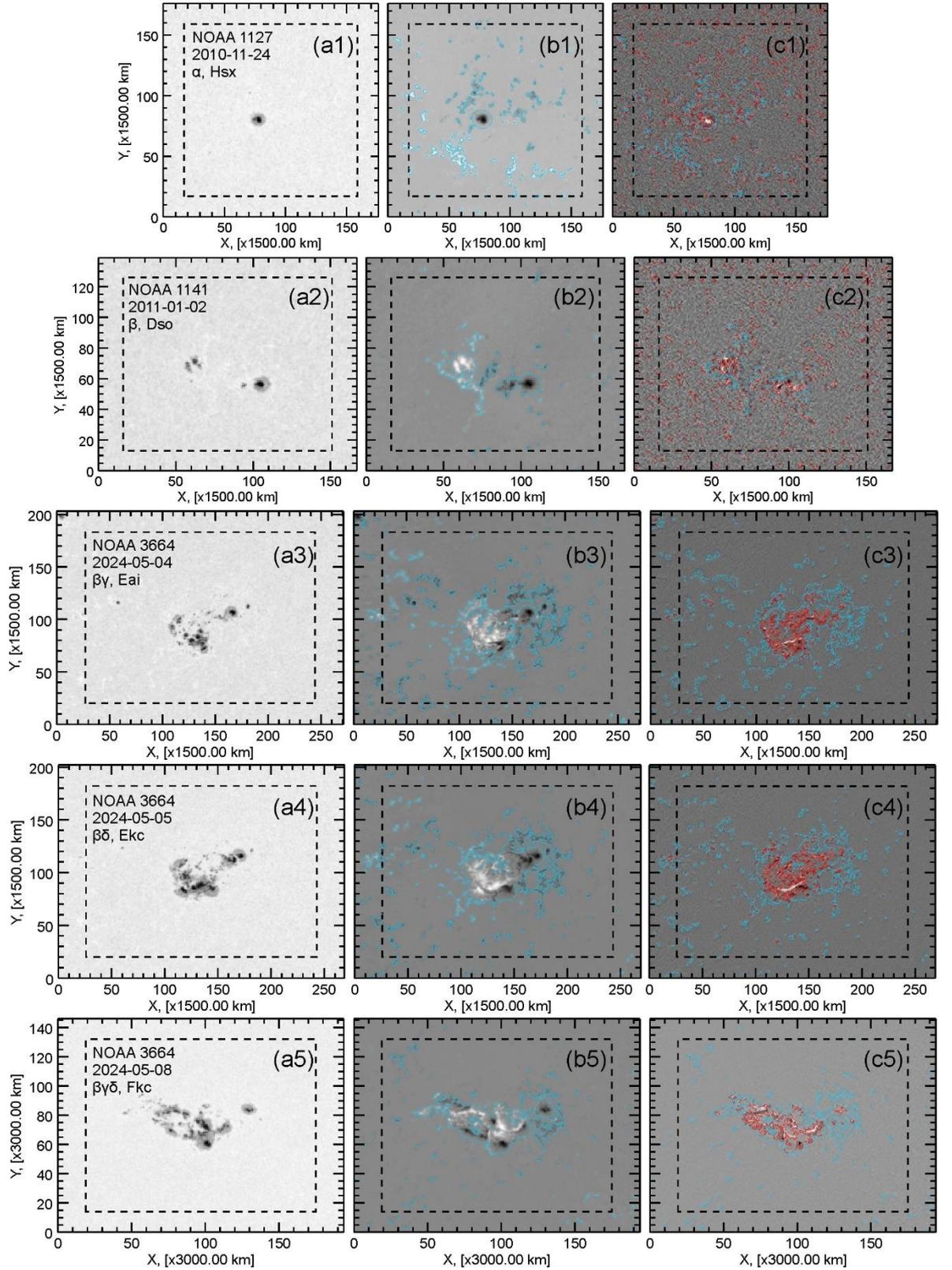

**Figure 1.** Examples of active regions of five Hale classes: (a1-c1) α, (a2-c2) β, (a3-c3) βγ, (a4-c4) βδ, (a5-c5) βγδ. (a) Distribution of intensity in the optical continuum, (b) vertical component of magnetic field $B_Z$ with iso-contours at $|B_Z|$=220 G, (c) vertical electric current density $j_Z$ with iso-contours at $|j_Z|$=9×10$^3$ statampere (red) and $|B_Z|$=220 G (cyan) (c) on the photosphere. Dashed rectangles show the projection of 3D boxes within which magnetic energies and other parameters were calculated.

**Figure 2.** Histograms with distributions of the number of analyzed active regions of Hale (a) and McIntosh (b) classes.

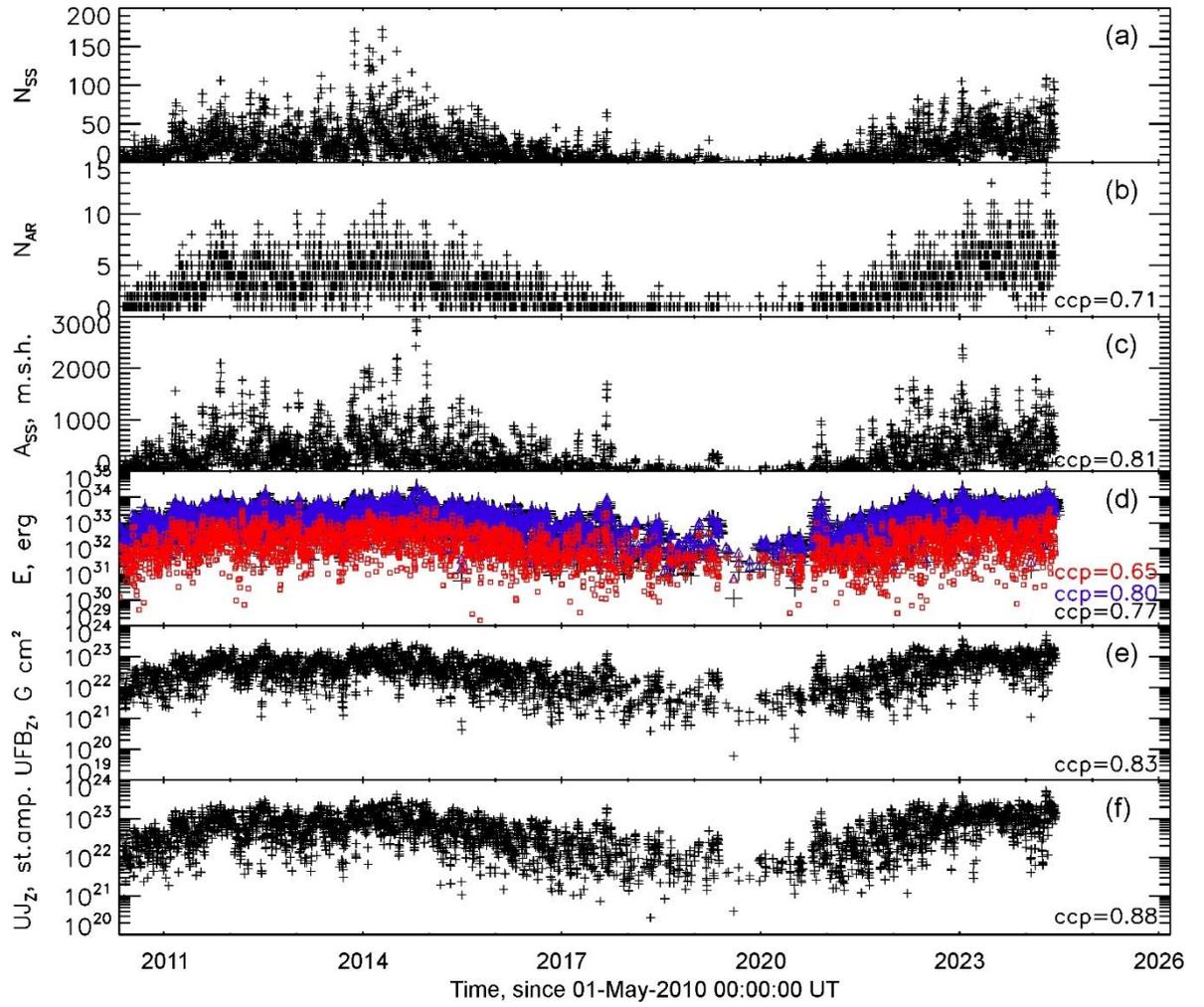

**Figure 3.** Temporal profiles of the following parameters integrated over ARs on the solar disk with a time step of 24 h: (a) number of sunspots, (b) number of active regions, (c) area of active regions, (d) magnetic energies ($E_{POTF}$ – black, $E_{NLFFF}$ – blue, $E_{FREE}$ – red), (e) unsigned magnetic flux, (f) unsigned vertical electric current on the photosphere.

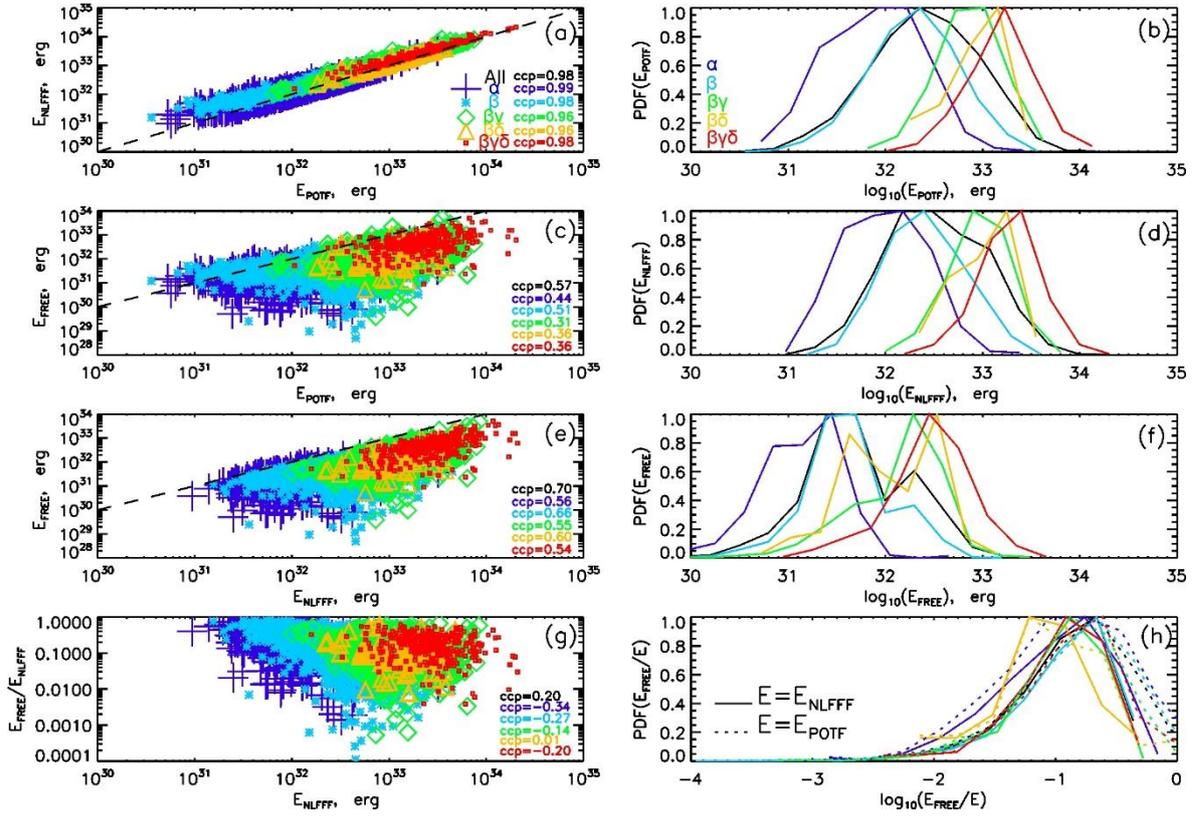

**Figure 4.** (*Left*) Pairwise dependences of magnetic energies on each other (a, c, e) and the fraction of the free magnetic energy on the nonlinear force-free field energy (g) for active regions of five Hale classes: α (blue cross), β (cyan asterisk), βγ (green diamond), βδ (orange triangle), βγδ (red square). The values of the Pearson correlation coefficients (*ccp*) are given in the lower right corners. The dependence *x=y* is shown by the straight dashed line in (a, c, e). (*Right*) Corresponding normalized distributions of the decimal logarithm of the magnetic energies (b, d, f) and the fraction of the free magnetic energy to the potential (dotted) and nonlinear force-free field (solid) energies (h).

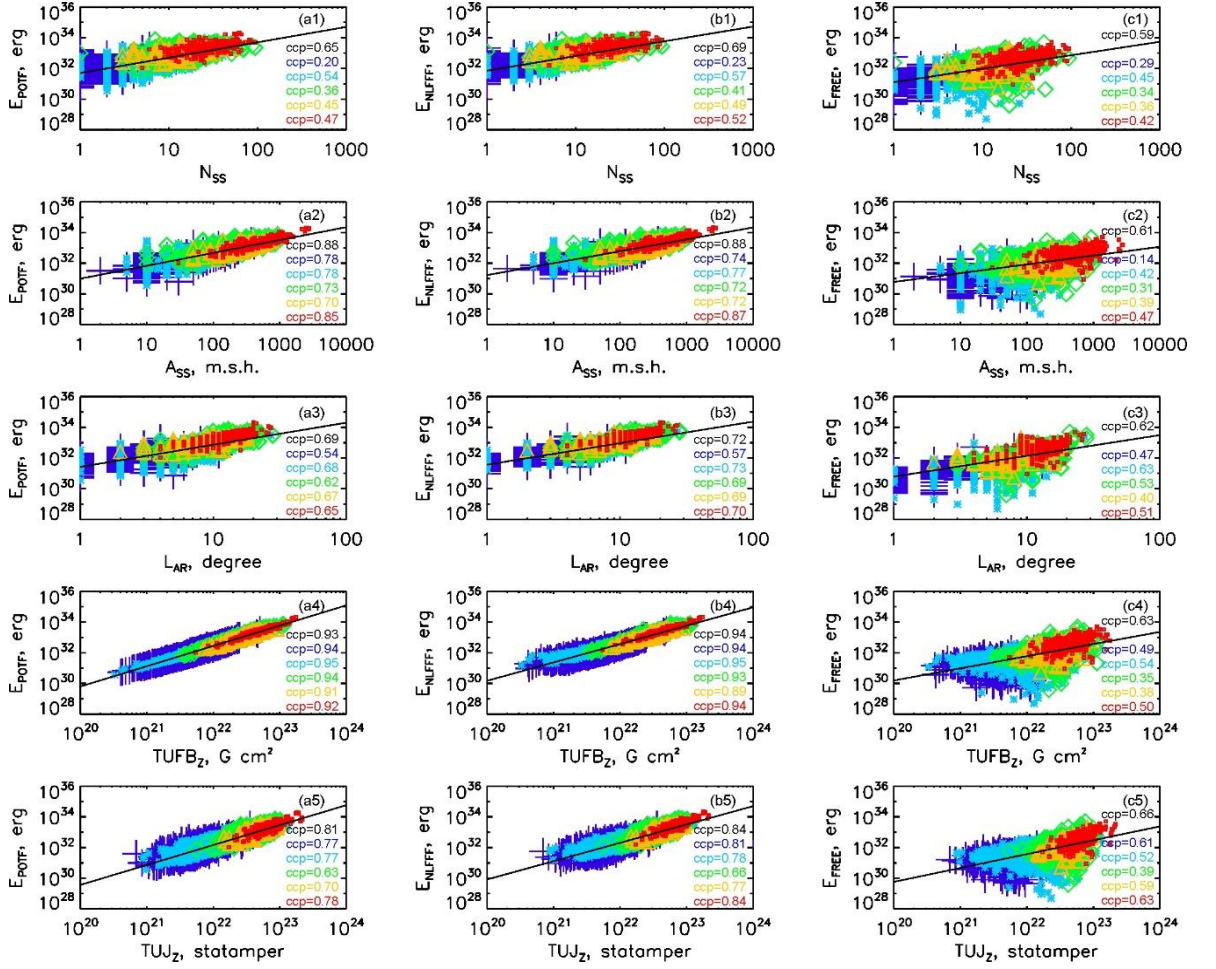

**Figure 5.** Pairwise dependences of magnetic energies, $E_{POTF}$ (left column, a1-a5), $E_{NLFFF}$ (middle column, b1-b5), $E_{FREE}$ (right column, c1-c5), on the following parameters: (a1-c1) number of sunspots, (a2-c2) area of sunspots, (a3-c3) longitudinal length of a sunspot group, (a4-c4) total unsigned magnetic flux, (a5-c5) total unsigned vertical electric current in active regions of five Hale classes (α - blue cross, β - cyan asterisk, βγ - green diamond, βδ - orange triangle, βγδ - red square). The values of the corresponding Pearson correlation coefficients (*ccp*) are given in the lower right corners. The straight black line shows the best fit of data points with the linear model in the log-log space.

**Figure 6.** Pairwise dependences of magnetic energies on each other (a, b, c) and the fraction of the free magnetic energy on the nonlinear force-free field energy (d) for active regions of McIntosh classes shown with different colors (see designations on the right). The dependence $x=y$ is shown by the straight dashed line in (a, b, c).

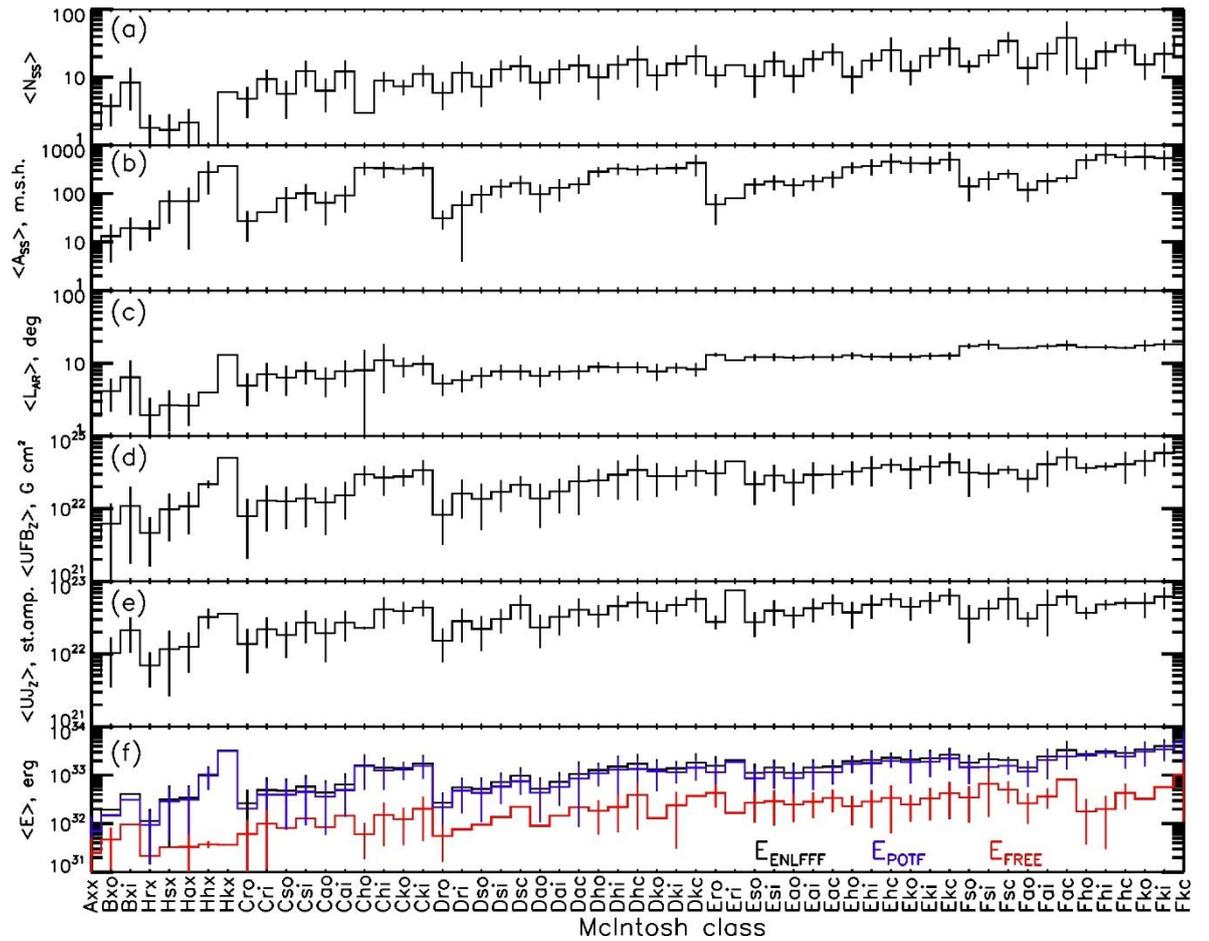

**Figure 7.** Average values of the following parameters of active regions of different McIntosh classes: (a) number of sunspots, (b) total area of sunspots, (c) longitudinal size of sunspot groups, (d) unsigned magnetic flux, (e) unsigned vertical electric current on the photosphere, (f) magnetic energies ($E_{NLFFF}$ – black, $E_{POTF}$ – blue, $E_{FREE}$ – red). Standard deviations are shown as vertical segments.

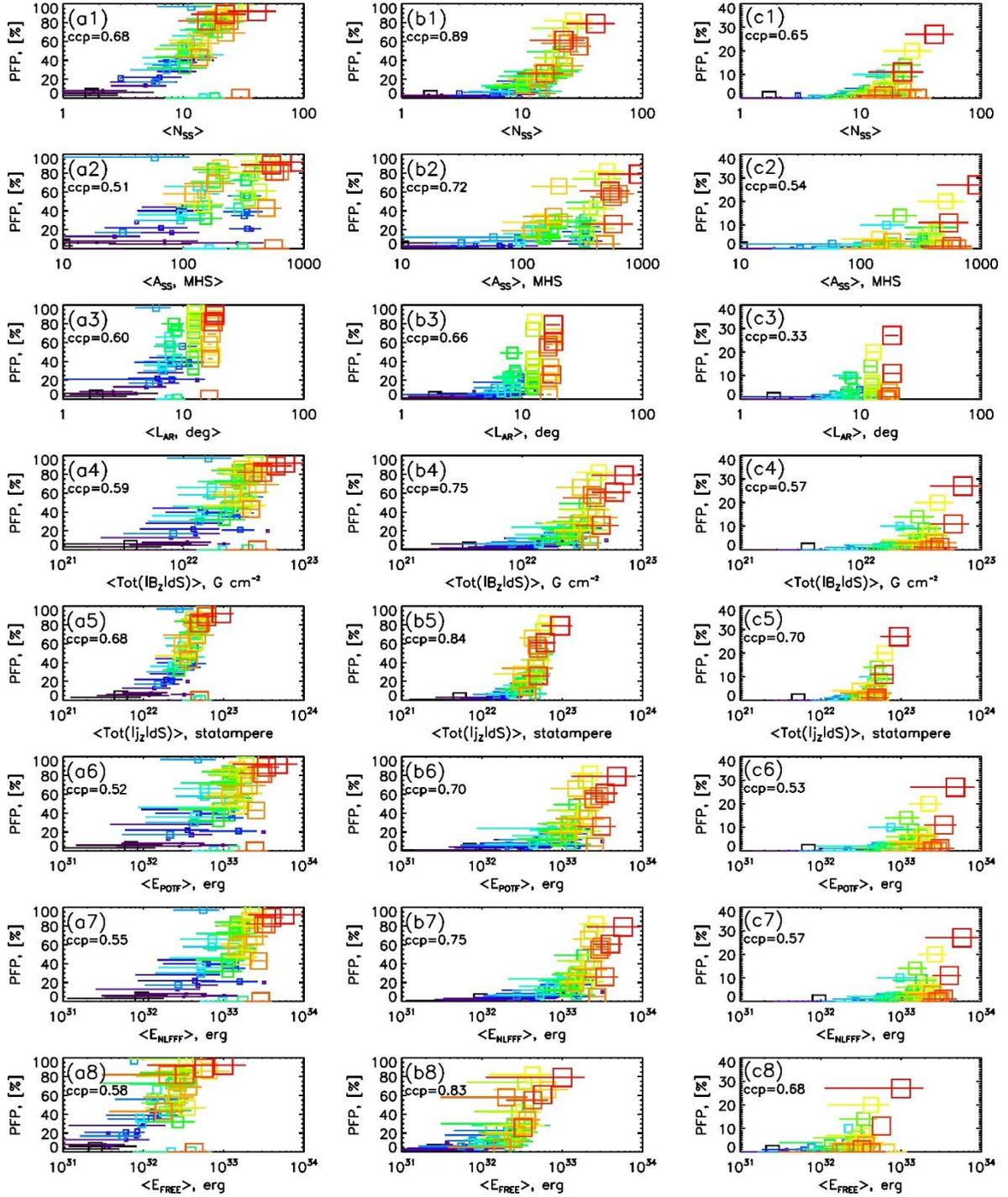

**Figure 8.** Pairwise dependences of Poisson flare probabilities of C (left column, a1-a8), M (middle column, b1-b8), and X (right column, c1-c8) class solar flares on the following parameters: (a1-c1) number of sunspots, (a2-c2) area of sunspots, (a3-b3) longitudinal size of sunspot group, (a4-c4) total unsigned magnetic flux, (a5-c5) total unsigned vertical electric current, (a6-c6) energy of potential magnetic field, (a7-c7) energy of nonlinear force-free field, (a8-c8) free magnetic energy, averaged over active regions of different McIntosh classes shown with squares of different size and color (from *Axx* – smallest black to *Fkc* – largest red). The standard deviations of the parameters from the mean are shown by horizontal segments of the corresponding color. The values of the Pearson correlation coefficients (*ccp*) are given in the upper left corners.

**Table 1.** Minimal (min), maximal (max), average (mean) of parameters, and standard deviations (sigma) for ARs of different Hale classes and all together, as well as their numbers and percentages.

| Parameter \ Hale AR | α | β | βγ | βδ | βγδ | All |
|---|---|---|---|---|---|---|
| Number of ARs | 509 | 3525 | 839 | 97 | 324 | 5294 |
| Percent. of ARs, % | 9.6 | 66.6 | 15.9 | 1.8 | 6.1 | 100 |
| Number of unique ARs | 351 | 1323 | 364 | 58 | 123 | 2219 |
| Percent. of unique ARs, % | 15.8 | 59.6 | 16.4 | 2.6 | 5.5 | 100 |
| min($N_{SS}$) | 1 | 1 | 1 | 3 | 5 | 1 |
| max($N_{SS}$) | 12 | 40 | 94 | 52 | 94 | 94 |
| mean($N_{SS}$) | 1.8 | 7.7 | 18.1 | 16.3 | 29.7 | 10.3 |
| sigma($N_{SS}$) | 1.2 | 5.5 | 9.8 | 9.5 | 15.8 | 9.8 |
| min($A_{SS}$), m.h.s. | 2 | 5 | 10 | 50 | 30 | 2 |
| max($A_{SS}$), m.h.s. | 410 | 1080 | 1100 | 770 | 2740 | 2740 |
| mean($A_{SS}$), m.h.s. | 41 | 90 | 269 | 328 | 609 | 151 |
| sigma($A_{SS}$), m.h.s. | 51 | 107 | 180 | 168 | 425 | 212 |
| min($L_{AR}$), deg | 1 | 1 | 3 | 2 | 4 | 1 |
| max($L_{AR}$), deg | 13 | 27 | 28 | 18 | 27 | 28 |
| mean($L_{AR}$), deg | 2.3 | 6.5 | 11.0 | 9.6 | 13.3 | 7.3 |
| sigma($L_{AR}$), deg | 1.6 | 3.2 | 3.4 | 3.8 | 3.9 | 4.2 |
| min(max($|B_Z|$)), G | 487 | 622 | 1008 | 1746 | 1664 | 487 |
| max(max($|B_Z|$)), G | 3347 | 3469 | 3661 | 3478 | 3623 | 3661 |
| mean(max($|B_Z|$)), G | 1790 | 2017 | 2436 | 2520 | 2644 | 2110 |
| sigma(max($|B_Z|$)), G | 546 | 442 | 370 | 352 | 393 | 496 |
| min(max($|B_H|$)), G | 244 | 291 | 485 | 849 | 1008 | 244 |
| max(max($|B_H|$)), G | 2219 | 2971 | 3371 | 3268 | 4166 | 4166 |
| mean(max($|B_H|$)), G | 919 | 1094 | 1416 | 1673 | 1812 | 1183 |
| sigma(max($|B_H|$)), G | 323 | 296 | 295 | 455 | 491 | 388 |
| min($\Sigma|B_Z|dS$), G cm$^2$ | $4.1\times10^{20}$ | $3.3\times10^{20}$ | $3.6\times10^{21}$ | $7.1\times10^{21}$ | $5.5\times10^{21}$ | $3.3\times10^{20}$ |
| max($\Sigma|B_Z|dS$), G cm$^2$ | $5.0\times10^{22}$ | $9.1\times10^{22}$ | $1.1\times10^{23}$ | $9.1\times10^{22}$ | $1.8\times10^{23}$ | $1.8\times10^{23}$ |
| mean($\Sigma|B_Z|dS$), G cm$^2$ | $7.0\times10^{21}$ | $1.3\times10^{22}$ | $3.0\times10^{22}$ | $3.2\times10^{22}$ | $5.0\times10^{22}$ | $1.8\times10^{22}$ |
| sigma($\Sigma|B_Z|dS$), G cm$^2$ | $6.1\times10^{21}$ | $1.1\times10^{22}$ | $1.7\times10^{22}$ | $1.5\times10^{22}$ | $2.8\times10^{22}$ | $1.7\times10^{22}$ |
| min($\Sigma|j_Z|dS$), statampere | $6.9\times10^{20}$ | $8.8\times10^{20}$ | $5.6\times10^{21}$ | $1.5\times10^{22}$ | $1.7\times10^{22}$ | $6.9\times10^{20}$ |
| max($\Sigma|j_Z|dS$), statampere | $7.2\times10^{22}$ | $1.3\times10^{23}$ | $1.0\times10^{23}$ | $9.6\times10^{22}$ | $2.2\times10^{23}$ | $2.2\times10^{23}$ |
| mean($\Sigma|j_Z|dS$), statampere | $8.8\times10^{21}$ | $2.1\times10^{22}$ | $4.5\times10^{22}$ | $4.8\times10^{22}$ | $7.2\times10^{22}$ | $2.7\times10^{22}$ |
| sigma($\Sigma|j_Z|dS$), statampere | $7.5\times10^{21}$ | $1.4\times10^{22}$ | $1.8\times10^{22}$ | $1.8\times10^{22}$ | $2.3\times10^{22}$ | $2.2\times10^{22}$ |
| min($E_{POTF}$), erg | $5.3\times10^{30}$ | $3.6\times10^{30}$ | $6.6\times10^{31}$ | $1.8\times10^{32}$ | $1.1\times10^{32}$ | $3.6\times10^{30}$ |
| max($E_{POTF}$), erg | $3.1\times10^{33}$ | $4.7\times10^{33}$ | $7.5\times10^{33}$ | $3.8\times10^{33}$ | $2.0\times10^{34}$ | $2.0\times10^{34}$ |
| mean($E_{POTF}$), erg | $1.9 \times 10^{32}$ | $4.5 \times 10^{32}$ | $1.3 \times 10^{33}$ | $1.4 \times 10^{33}$ | $2.8 \times 10^{33}$ | $7.4 \times 10^{32}$ |
| sigma($E_{POTF}$), erg | $2.5 \times 10^{32}$ | $5.2 \times 10^{32}$ | $1.0 \times 10^{33}$ | $8.5 \times 10^{32}$ | $2.6 \times 10^{33}$ | $1.1 \times 10^{33}$ |
| min($E_{NLFFF}$), erg | $9.5 \times 10^{30}$ | $1.6 \times 10^{31}$ | $1.0 \times 10^{32}$ | $2.2 \times 10^{32}$ | $1.6 \times 10^{32}$ | $9.5 \times 10^{30}$ |
| max($E_{NLFFF}$), erg | $3.2 \times 10^{33}$ | $5.5 \times 10^{33}$ | $8.4 \times 10^{33}$ | $4.8 \times 10^{33}$ | $2.1 \times 10^{34}$ | $2.1 \times 10^{34}$ |
| mean($E_{NLFFF}$), erg | $2.2 \times 10^{32}$ | $5.5 \times 10^{32}$ | $1.6 \times 10^{33}$ | $1.7 \times 10^{33}$ | $3.4 \times 10^{33}$ | $8.9 \times 10^{32}$ |
| sigma($E_{NLFFF}$), erg | $2.7 \times 10^{32}$ | $5.9 \times 10^{32}$ | $1.2 \times 10^{33}$ | $9.9 \times 10^{32}$ | $2.9 \times 10^{33}$ | $1.3 \times 10^{33}$ |
| min($E_{FREE}$), erg | $4.4 \times 10^{29}$ | $5.0 \times 10^{28}$ | $3.8 \times 10^{29}$ | $5.4 \times 10^{30}$ | $8.9 \times 10^{30}$ | $5.0 \times 10^{28}$ |
| max($E_{FREE}$), erg | $5.7 \times 10^{32}$ | $2.5 \times 10^{33}$ | $4.9 \times 10^{33}$ | $2.1 \times 10^{33}$ | $5.0 \times 10^{33}$ | $5.0 \times 10^{33}$ |
| mean($E_{FREE}$), erg | $2.8 \times 10^{31}$ | $9.9 \times 10^{31}$ | $2.9 \times 10^{32}$ | $2.6 \times 10^{32}$ | $6.1 \times 10^{32}$ | $1.6 \times 10^{32}$ |
| sigma($E_{FREE}$), erg | $4.0 \times 10^{31}$ | $1.3 \times 10^{32}$ | $3.2 \times 10^{32}$ | $2.9 \times 10^{32}$ | $6.2 \times 10^{32}$ | $2.7 \times 10^{32}$ |
| min($E_{FREE}/E_{NLFFF}$) | $1.4 \times 10^{-3}$ | $1.1 \times 10^{-4}$ | $5.3 \times 10^{-4}$ | $7.7 \times 10^{-3}$ | $3.9 \times 10^{-3}$ | $1.1 \times 10^{-4}$ |
| max($E_{FREE}/E_{NLFFF}$) | 0.79 | 0.86 | 0.67 | 0.60 | 0.62 | 0.86 |
| mean($E_{FREE}/E_{NLFFF}$) | 0.20 | 0.23 | 0.19 | 0.15 | 0.20 | 0.22 |
| sigma($E_{FREE}/E_{NLFFF}$) | 0.17 | 0.15 | 0.12 | 0.12 | 0.13 | 0.15 |
| min($E_{FREE}/E_{POTF}$) | $1.4 \times 10^{-3}$ | $1.1 \times 10^{-4}$ | $5.3 \times 10^{-4}$ | $7.7 \times 10^{-3}$ | $3.9 \times 10^{-3}$ | $1.1 \times 10^{-4}$ |
| max($E_{FREE}/E_{POTF}$) | 3.83 | 6.18 | 2.04 | 1.51 | 1.63 | 6.18 |
| mean($E_{FREE}/E_{POTF}$) | 0.34 | 0.37 | 0.27 | 0.22 | 0.29 | 0.34 |
| sigma($E_{FREE}/E_{POTF}$) | 0.45 | 0.41 | 0.24 | 0.25 | 0.25 | 0.38 |

**Table 2.** Values of the coefficients *a, b* and of the unreduced chi-square goodness-of-fit statistic of the power-law model (5) used to approximate the dependences of the magnetic energies on the considered ARs' physical parameters.

| Parameter | $N_{SS}$ | $A_{SS}$ | $L_{AR}$ | $TUFB_Z$ | $TUJ_Z$ |
|---|---|---|---|---|---|
| $E_{POTF}$ | a=31.70±0.97<br>b=1.00±1.06<br>$\chi^2$=0.83 | a=30.99±1.44<br>b=0.84±0.75<br>$\chi^2$=0.48 | a=31.42±1.18<br>b=1.44±1.43<br>$\chi^2$=0.71 | a=3.32±22.45<br>b=1.32±1.02<br>$\chi^2$=0.04 | a=3.56±0.97<br>b=1.30±1.06<br>$\chi^2$=0.36 |
| $E_{NLFFF}$ | a=31.85±1.00<br>b=0.96±1.09<br>$\chi^2$=0.60 | a=31.22±1.48<br>b=0.78±0.77<br>$\chi^2$=0.36 | a=31.57±1.21<br>b=1.41±1.47<br>$\chi^2$=0.47 | a=6.21±23.10<br>b=1.20±1.05<br>$\chi^2$=0.07 | a=5.98±0.97<br>b=1.20±1.06<br>$\chi^2$=0.28 |
| $E_{FREE}$ | a=31.10±0.99<br>b=0.88±1.08<br>$\chi^2$=1.11 | a=30.78±1.48<br>b=0.57±0.77<br>$\chi^2$=1.22 | a=30.79±1.21<br>b=1.35±1.46<br>$\chi^2$=0.91 | a=14.36±23.06<br>b=0.79±1.04<br>$\chi^2$=1.20 | a=11.68±0.97<br>b=0.90±1.06<br>$\chi^2$=1.14 |

**Table 3.** Minimal (min), maximal (max), average (mean) of parameters, and standard deviations (sigma) for ARs of different McIntosh classes, as well as their numbers and percentages.

| AR param.<br>McIntosh class | Number of ARs / Number of unique ARs | Percentage of ARs / unique ARs, % | $N_{SS}$<br>min<br>max<br>mean<br>sigma | $A_{SS}$, m.h.s.<br>min<br>max<br>mean<br>sigma | $L_{AR}$, deg<br>min<br>max<br>mean<br>sigma | $max(|B_Z|)$, G<br>min<br>max<br>mean<br>sigma | $max(|B_H|)$, G<br>min<br>max<br>mean<br>sigma | $\Sigma|B_Z|dS$, G cm$^2$<br>min<br>max<br>mean<br>sigma | $\Sigma|j_Z|dS$, statamp.<br>min<br>max<br>mean<br>sigma | $E_{POTF}$, erg<br>min<br>max<br>mean<br>sigma | $E_{NLFFF}$, erg<br>min<br>max<br>mean<br>sigma | $E_{FREE}$, erg<br>min<br>max<br>mean<br>sigma | $E_{FREE}/E_{POTF}$<br>min<br>max<br>mean<br>sigma | $E_{FREE}/E_{NLFFF}$<br>min<br>max<br>mean<br>sigma |
|---|---|---|---|---|---|---|---|---|---|---|---|---|---|---|
| 1 | 2 | 3 | 4 | 5 | 6 | 7 | 8 | 9 | 10 | 11 | 12 | 13 | 14 | 15 |
| All | 5288 / 4181 | 100.00 / 100.00 | 1.00<br>94.00<br>13.96<br>8.88 | 2.00<br>2740.00<br>242.93<br>190.55 | 1.00<br>28.00<br>10.29<br>4.63 | 487.32<br>3660.87<br>2383.20<br>400.24 | 243.50<br>4166.29<br>1355.71<br>262.92 | 3.30×10$^{20}$<br>1.76×10$^{23}$<br>2.72×10$^{22}$<br>1.40×10$^{22}$ | 6.91×10$^{20}$<br>2.17×10$^{23}$<br>3.74×10$^{22}$<br>1.74×10$^{22}$ | 3.60×10$^{30}$<br>2.03×10$^{34}$<br>1.30×10$^{33}$<br>9.55×10$^{32}$ | 9.47×10$^{30}$<br>2.09×10$^{34}$<br>1.54×10$^{33}$<br>1.10×10$^{33}$ | 4.97×10$^{28}$<br>5.02×10$^{33}$<br>2.39×10$^{32}$<br>1.95×10$^{32}$ | 1.09×10$^{-4}$<br>6.18×10$^{0}$<br>2.63×10$^{-1}$<br>1.32×10$^{-1}$ | 1.09×10$^{-4}$<br>8.61×10$^{-1}$<br>1.79×10$^{-1}$<br>7.02×10$^{-2}$ |
| Axx | 212 / 173 | 4.01 / 4.14 | 1.00<br>12.00<br>1.73<br>1.11 | 2.00<br>30.00<br>10.03<br>1.93 | 1.00<br>11.00<br>1.89<br>1.48 | 487.32<br>2501.84<br>1342.66<br>341.56 | 243.50<br>1367.62<br>652.87<br>199.80 | 4.12×10$^{20}$<br>1.63×10$^{22}$<br>3.63×10$^{21}$<br>2.87×10$^{21}$ | 6.91×10$^{20}$<br>2.66×10$^{22}$<br>5.25×10$^{21}$<br>4.00×10$^{21}$ | 5.28×10$^{30}$<br>4.45×10$^{32}$<br>7.03×10$^{31}$<br>7.00×10$^{31}$ | 9.47×10$^{30}$<br>5.00×10$^{32}$<br>9.54×10$^{31}$<br>8.42×10$^{31}$ | 5.15×10$^{29}$<br>1.92×10$^{32}$<br>2.52×10$^{31}$<br>2.51×10$^{31}$ | 5.56×10$^{-3}$<br>3.83×10$^{0}$<br>5.66×10$^{-1}$<br>5.79×10$^{-1}$ | 5.52×10$^{-3}$<br>7.93×10$^{-1}$<br>3.01×10$^{-1}$<br>1.80×10$^{-1}$ |
| Bxo | 678 / 576 | 12.82 / 13.78 | 1.00<br>15.00<br>3.78<br>1.88 | 5.00<br>160.00<br>13.33<br>9.48 | 1.00<br>14.00<br>4.12<br>1.97 | 639.44<br>2594.21<br>1559.65<br>342.08 | 349.09<br>1623.10<br>799.13<br>230.98 | 3.30×10$^{20}$<br>6.61×10$^{22}$<br>6.20×10$^{21}$<br>5.50×10$^{21}$ | 9.28×10$^{20}$<br>7.34×10$^{22}$<br>1.02×10$^{21}$<br>6.78×10$^{21}$ | 3.60×10$^{30}$<br>2.89×10$^{33}$<br>1.48×10$^{32}$<br>1.88×10$^{32}$ | 1.57×10$^{31}$<br>2.98×10$^{33}$<br>1.95×10$^{32}$<br>1.99×10$^{32}$ | 9.44×10$^{28}$<br>3.65×10$^{32}$<br>4.68×10$^{31}$<br>3.50×10$^{31}$ | 6.25×10$^{-4}$<br>6.18×10$^{0}$<br>6.10×10$^{-1}$<br>6.38×10$^{-1}$ | 6.25×10$^{-4}$<br>8.61×10$^{-1}$<br>3.16×10$^{-1}$<br>1.81×10$^{-1}$ |
| Bxi | 13 / 10 | 0.25 / 0.24 | 2.00<br>21.00<br>8.38<br>5.16 | 10.00<br>50.00<br>19.23<br>12.56 | 3.00<br>19.00<br>6.38<br>4.43 | 1064.96<br>2807.62<br>1769.88<br>473.42 | 640.12<br>1701.78<br>983.22<br>287.30 | 1.72×10$^{21}$<br>3.83×10$^{22}$<br>1.08×10$^{22}$<br>9.07×10$^{21}$ | 5.18×10$^{21}$<br>4.51×10$^{22}$<br>2.11×10$^{22}$<br>1.06×10$^{22}$ | 2.94×10$^{31}$<br>1.50×10$^{33}$<br>3.13×10$^{32}$<br>3.73×10$^{32}$ | 6.77×10$^{31}$<br>2.04×10$^{33}$<br>4.10×10$^{32}$<br>5.04×10$^{32}$ | 2.32×10$^{31}$<br>5.42×10$^{32}$<br>9.70×10$^{31}$<br>1.39×10$^{32}$ | 5.32×10$^{-2}$<br>1.20×10$^{0}$<br>3.87×10$^{-1}$<br>3.17×10$^{-1}$ | 5.05×10$^{-2}$<br>5.65×10$^{-1}$<br>2.52×10$^{-1}$<br>1.33×10$^{-1}$ |
| Hrx | 49 / 32 | 0.93 / 0.77 | 1.00<br>6.00<br>1.80<br>1.00 | 10.00<br>40.00<br>18.98<br>8.48 | 1.00<br>7.00<br>1.92<br>1.41 | 1099.29<br>2136.90<br>1657.97<br>258.95 | 505.95<br>1160.21<>852.66<br>144.88 | 1.01×10$^{21}$<br>1.69×10$^{22}$<br>4.64×10$^{21}$<br>3.03×10$^{21}$ | 1.64×10$^{21}$<br>1.71×10$^{22}$<br>7.00×10$^{21}$<br>3.51×10$^{21}$ | 1.60×10$^{31}$<br>4.73×10$^{32}$<br>9.35×10$^{31}$<br>7.90×10$^{31}$ | 2.32×10$^{31}$<br>5.08×10$^{32}$<br>1.15×10$^{32}$<br>8.49×10$^{31}$ | 5.38×10$^{29}$<br>9.63×10$^{31}$<br>2.17×10$^{31}$<br>2.29×10$^{31}$ | 2.00×10$^{-3}$<br>1.08×10$^{0}$<br>3.11×10$^{-1}$<br>2.83×10$^{-1}$ | 2.00×10$^{-3}$<br>5.19×10$^{-1}$<br>2.06×10$^{-1}$<br>1.53×10$^{-1}$ |
| Hsx | 167 / 137 | 3.16 / 3.28 | 1.00<br>6.00<br>1.67<br>1.16 | 10.00<br>240.00<br>70.18<br>46.13 | 1.00<br>8.00<br>2.66<br>1.52 | 825.70<br>3129.75<br>2251.77<br>372.48 | 511.05<br>2219.37<br>1178.19<br>224.16 | 8.16×10$^{20}$<br>4.67×10$^{22}$<br>9.79×10$^{21}$<br>6.24×10$^{21}$ | 2.22×10$^{21}$<br>7.19×10$^{22}$<br>1.17×10$^{22}$<br>9.07×10$^{21}$ | 1.21×10$^{31}$<br>1.94×10$^{33}$<br>2.84×10$^{32}$<br>2.39×10$^{32}$ | 2.87×10$^{31}$<br>2.46×10$^{33}$<br>3.17×10$^{32}$<br>2.84×10$^{32}$ | 4.44×10$^{29}$<br>5.66×10$^{32}$<br>3.27×10$^{31}$<br>5.94×10$^{31}$ | 1.38×10$^{-3}$<br>1.43×10$^{0}$<br>1.48×10$^{-1}$<br>1.18×10$^{-1}$ | 1.38×10$^{-3}$<br>5.89×10$^{-1}$<br>1.15×10$^{-1}$<br>9.45×10$^{-2}$ |
| Hax | 77 / 57 | 1.46 / 1.36 | 1.00<br>7.00<br>2.13<br>1.24 | 10.00<br>310.00<br>69.48<br>62.30 | 1.00<br>6.00<br>2.60<br>1.23 | 657.80<br>3336.41<br>2058.28<br>381.15 | 318.67<br>1861.48<br>1101.07<br>242.76 | 4.79×10$^{20}$<br>2.49×10$^{22}$<br>1.07×10$^{22}$<br>6.24×10$^{21}$ | 1.55×10$^{21}$<br>3.78×10$^{22}$<br>1.26×10$^{22}$<br>7.06×10$^{21}$ | 6.36×10$^{30}$<br>1.35×10$^{33}$<br>3.11×10$^{32}$<br>2.58×10$^{32}$ | 1.40×10$^{31}$<br>1.37×10$^{33}$<br>3.44×10$^{32}$<br>2.66×10$^{32}$ | 7.34×10$^{29}$<br>1.04×10$^{32}$<br>3.32×10$^{31}$<br>2.54×10$^{31}$ | 3.16×10$^{-3}$<br>1.31×10$^{0}$<br>1.83×10$^{-1}$<br>2.30×10$^{-1}$ | 3.15×10$^{-3}$<br>5.68×10$^{-1}$<br>1.33×10$^{-1}$<br>1.17×10$^{-1}$ |
| Hhx | 2 / 2 | 0.04 / 0.05 | 1.00<br>1.00<br>1.00<br>0.00 | 150.00<br>410.00<br>280.00<br>183.85 | 4.00<br>4.00<br>4.00<br>0.00 | 2766.13<br>3346.63<br>3056.38<br>410.48 | 1524.82<br>1807.35<br>1666.09<br>199.78 | 2.00×10$^{22}$<br>2.35×10$^{22}$<br>2.17×10$^{22}$<br>2.48×10$^{21}$ | 2.52×10$^{22}$<br>3.93×10$^{22}$<br>3.22×10$^{22}$<br>1.00×10$^{22}$ | 6.36×10$^{32}$<br>1.32×10$^{33}$<br>9.80×10$^{32}$<br>4.86×10$^{32}$ | 6.69×10$^{32}$<br>1.36×10$^{33}$<br>1.02×10$^{33}$<br>4.92×10$^{32}$ | 3.32×10$^{31}$<br>4.18×10$^{31}$<br>3.75×10$^{31}$<br>6.11×10$^{30}$ | 3.16×10$^{-2}$<br>5.21×10$^{-2}$<br>4.19×10$^{-2}$<br>1.45×10$^{-2}$ | 3.06×10$^{-2}$<br>4.95×10$^{-2}$<br>4.01×10$^{-2}$<br>1.34×10$^{-2}$ |
| Hkx | 1 / | 0.02 / | 6.00 | 370.00 | 13.00 | 3057.27 | 1652.83 | 5.00×10$^{22}$ | 3.58×10$^{22}$ | 3.15×10$^{33}$ | 3.18×10$^{33}$ | 3.66×10$^{31}$ | 1.16×10$^{-2}$ | 1.15×10$^{-2}$ |

| | | | | | | | | | | | | | |
|---|---|---|---|---|---|---|---|---|---|---|---|---|---|
| | | 1 | 0.02 | 6.00 | 370.00 | 13.00 | 3057.27 | 1652.83 | 5.00×10²² | 3.58×10²² | 3.15×10³³ | 3.18×10³³ | 3.66×10³¹ | 1.16×10⁻² | 1.15×10⁻² |
| | | | | 6.00 | 370.00 | 13.00 | 3057.27 | 1652.83 | 5.00×10²² | 3.58×10²² | 3.15×10³³ | 3.18×10³³ | 3.66×10³¹ | 1.16×10⁻² | 1.15×10⁻² |
| | | | | - | - | - | - | - | - | - | - | - | - | - | - |
| Cro | 456 / 376 | 8.62 / 8.99 | 1.00 | 10.00 | 1.00 | 865.75 | 437.89 | 1.07×10²¹ | 1.86×10²¹ | 1.68×10³¹ | 4.63×10³¹ | 1.75×10³⁰ | 5.39×10⁻³ | 5.36×10⁻³ |
| | | | 20.00 | 290.00 | 15.00 | 2973.06 | 1885.87 | 4.59×10²² | 5.20×10²² | 1.88×10³³ | 2.08×10³³ | 4.20×10³² | 2.49×10⁰ | 7.14×10⁻¹ |
| | | | 4.84 | 26.80 | 4.91 | 1783.43 | 946.35 | 7.87×10²¹ | 1.37×10²² | 2.04×10³² | 2.66×10³² | 6.12×10³¹ | 4.63×10⁻¹ | 2.78×10⁻¹ |
| | | | 2.33 | 16.50 | 2.30 | 296.11 | 203.33 | 5.82×10²¹ | 8.27×10²¹ | 2.07×10³² | 2.40×10³² | 5.78×10³¹ | 3.97×10⁻¹ | 1.52×10⁻¹ |
| Cri | 25 / 14 | 0.47 / 0.33 | 3.00 | 10.00 | 2.00 | 1302.57 | 649.36 | 4.59×10²¹ | 9.29×10²¹ | 9.15×10³¹ | 1.66×10³² | 5.34×10³⁰ | 1.75×10⁻² | 1.72×10⁻² |
| | | | 17.00 | 210.00 | 14.00 | 3004.59 | 1753.60 | 3.44×10²² | 6.02×10²² | 1.50×10³³ | 1.63×10³³ | 3.56×10³² | 1.24×10⁰ | 5.54×10⁻¹ |
| | | | 9.44 | 40.80 | 7.08 | 1893.92 | 1072.22 | 1.29×10²² | 2.21×10²² | 3.94×10³² | 4.93×10³² | 9.95×10³¹ | 3.38×10⁻¹ | 2.27×10⁻¹ |
| | | | 3.38 | 41.02 | 2.93 | 367.85 | 249.57 | 8.04×10²¹ | 9.99×10²¹ | 3.55×10³² | 4.12×10³² | 8.77×10³¹ | 2.73×10⁻¹ | 1.32×10⁻¹ |
| Cso | 369 / 299 | 6.98 / 7.15 | 1.00 | 10.00 | 1.00 | 1078.79 | 525.43 | 9.39×10²⁰ | 2.53×10²¹ | 1.55×10³¹ | 3.83×10³¹ | 1.98×10³⁰ | 2.97×10⁻³ | 2.96×10⁻³ |
| | | | 23.00 | 250.00 | 17.00 | 3404.50 | 1984.96 | 4.55×10²² | 6.81×10²² | 2.44×10³³ | 2.52×10³³ | 5.52×10³² | 2.11×10⁰ | 6.78×10⁻¹ |
| | | | 5.65 | 80.11 | 6.37 | 2181.17 | 1154.99 | 1.26×10²² | 1.82×10²² | 3.97×10³² | 4.79×10³² | 8.18×10³¹ | 2.78×10⁻¹ | 1.91×10⁻¹ |
| | | | 3.18 | 54.54 | 2.82 | 387.48 | 231.63 | 7.32×10²¹ | 9.41×10²¹ | 3.23×10³² | 3.66×10³² | 9.43×10³¹ | 2.74×10⁻¹ | 1.32×10⁻¹ |
| Csi | 27 / 14 | 0.51 / 0.33 | 5.00 | 30.00 | 3.00 | 1496.45 | 748.78 | 4.56×10²¹ | 1.02×10²² | 8.68×10³¹ | 1.58×10³² | 5.18×10²⁹ | 9.18×10⁻⁴ | 9.17×10⁻⁴ |
| | | | 24.00 | 240.00 | 12.00 | 3047.57 | 2088.19 | 4.62×10²² | 6.25×10²² | 1.95×10³³ | 2.13×10³³ | 4.26×10³² | 1.02×10⁰ | 5.04×10⁻¹ |
| | | | 12.26 | 101.11 | 7.81 | 2209.99 | 1220.20 | 1.38×10²² | 2.71×10²² | 4.53×10³² | 5.82×10³² | 1.29×10³² | 3.23×10⁻¹ | 2.14×10⁻¹ |
| | | | 5.02 | 56.11 | 2.63 | 387.88 | 275.86 | 8.26×10²¹ | 1.30×10²² | 3.59×10³² | 4.30×10³² | 1.32×10³² | 2.87×10⁻¹ | 1.48×10⁻¹ |
| Cao | 373 / 303 | 7.05 / 7.25 | 1.00 | 10.00 | 1.00 | 1103.38 | 551.36 | 9.80×10²⁰ | 3.77×10²¹ | 1.73×10³¹ | 7.93×10³¹ | 8.31×10²⁹ | 1.23×10⁻³ | 1.23×10⁻³ |
| | | | 23.00 | 240.00 | 15.00 | 3153.75 | 2567.66 | 4.92×10²² | 1.22×10²³ | 2.16×10³³ | 2.73×10³³ | 8.88×10³² | 3.59×10⁰ | 7.82×10⁻¹ |
| | | | 6.32 | 65.05 | 6.09 | 2045.54 | 1120.53 | 1.21×10²² | 1.92×10²² | 3.56×10³² | 4.39×10³² | 8.38×10³¹ | 3.21×10⁻¹ | 2.11×10⁻¹ |
| | | | 3.26 | 42.87 | 2.64 | 313.81 | 234.87 | 7.70×10²¹ | 1.15×10²² | 2.97×10³² | 3.53×10³² | 9.76×10³¹ | 3.32×10⁻¹ | 1.40×10⁻¹ |
| Cai | 61 / 42 | 1.15 / 1.00 | 4.00 | 10.00 | 3.00 | 1318.98 | 701.68 | 2.25×10²¹ | 6.83×10²¹ | 3.85×10³¹ | 8.59×10³¹ | 8.23×10³⁰ | 1.31×10⁻² | 1.30×10⁻² |
| | | | 29.00 | 220.00 | 19.00 | 2910.40 | 1837.62 | 3.59×10²² | 7.23×10²² | 1.48×10³³ | 2.07×10³³ | 6.57×10³² | 1.23×10⁰ | 5.51×10⁻¹ |
| | | | 12.23 | 91.72 | 7.72 | 2016.39 | 1125.38 | 1.51×10²² | 2.71×10²² | 4.93×10³² | 6.39×10³² | 1.46×10³² | 3.25×10⁻¹ | 2.22×10⁻¹ |
| | | | 5.47 | 50.19 | 3.02 | 314.64 | 217.51 | 8.05×10²¹ | 1.21×10²² | 3.53×10³² | 4.63×10³² | 1.58×10³² | 2.55×10⁻¹ | 1.28×10⁻¹ |
| Cho | 2 / 2 | 0.04 / 0.05 | 3.00 | 270.00 | 3.00 | 2445.68 | 1397.04 | 2.35×10²² | 2.22×10²² | 8.06×10³² | 8.37×10³² | 3.09×10³¹ | 3.83×10⁻² | 3.69×10⁻² |
| | | | 3.00 | 410.00 | 13.00 | 2994.15 | 1580.70 | 3.59×10²² | 2.34×10²² | 2.26×10³³ | 2.35×10³³ | 9.04×10³¹ | 4.00×10⁻² | 3.85×10⁻² |
| | | | 3.00 | 340.00 | 8.00 | 2719.92 | 1488.87 | 2.97×10²² | 2.28×10²² | 1.53×10³³ | 1.59×10³³ | 6.07×10³¹ | 3.92×10⁻² | 3.77×10⁻² |
| | | | 0.00 | 98.99 | 7.07 | 387.83 | 129.87 | 8.83×10²¹ | 8.12×10²⁰ | 1.03×10³³ | 1.07×10³³ | 4.21×10³¹ | 1.20×10⁻³ | 1.11×10⁻³ |
| Chi | 2 / 2 | 0.04 / 0.05 | 7.00 | 250.00 | 6.00 | 2847.78 | 1490.20 | 1.86×10²² | 2.79×10²² | 7.25×10³² | 7.94×10³² | 6.87×10³¹ | 9.48×10⁻² | 8.66×10⁻² |
| | | | 11.00 | 420.00 | 16.00 | 2886.03 | 1630.50 | 3.49×10²² | 5.43×10²² | 1.80×10³³ | 2.04×10³³ | 2.35×10³² | 1.30×10⁻¹ | 1.15×10⁻¹ |
| | | | 9.00 | 335.00 | 11.00 | 2866.91 | 1560.35 | 2.67×10²² | 4.11×10²² | 1.26×10³³ | 1.42×10³³ | 1.52×10³² | 1.12×10⁻¹ | 1.01×10⁻¹ |
| | | | 2.83 | 120.21 | 7.07 | 27.04 | 99.21 | 1.15×10²² | 1.87×10²² | 7.63×10³² | 8.80×10³² | 1.17×10³² | 2.50×10⁻² | 2.02×10⁻² |
| Cko | 11 / 8 | 0.21 / 0.19 | 4.00 | 250.00 | 4.00 | 2464.68 | 1189.74 | 1.64×10²² | 1.53×10²² | 6.69×10³² | 7.73×10³² | 9.95×10³⁰ | 7.72×10⁻³ | 7.66×10⁻³ |
| | | | 10.00 | 460.00 | 12.00 | 2938.12 | 2550.78 | 4.29×10²² | 5.30×10²² | 2.43×10³³ | 2.51×10³³ | 2.56×10³² | 2.52×10⁻¹ | 2.02×10⁻¹ |
| | | | 7.36 | 319.09 | 9.18 | 2658.15 | 1580.12 | 2.77×10²² | 3.90×10²² | 1.28×10³³ | 1.40×10³³ | 1.25×10³² | 1.10×10⁻¹ | 9.44×10⁻² |
| | | | 1.86 | 68.77 | 2.75 | 166.13 | 357.26 | 7.10×10²² | 1.31×10²² | 4.69×10³² | 4.64×10³² | 8.88×10³¹ | 8.43×10⁻² | 6.71×10⁻² |
| Cki | 5 / 4 | 0.09 / 0.10 | 6.00 | 230.00 | 6.00 | 2625.70 | 1435.74 | 2.52×10²² | 3.34×10²² | 9.81×10³² | 1.04×10³³ | 4.88×10³¹ | 3.60×10⁻³ | 3.59×10⁻³ |
| | | | 16.00 | 480.00 | 14.00 | 2793.38 | 1679.26 | 5.50×10²² | 6.00×10²² | 2.78×10³³ | 3.14×10³³ | 3.60×10³² | 2.45×10⁻¹ | 1.97×10⁻¹ |
| | | | 11.20 | 336.00 | 9.80 | 2700.39 | 1577.53 | 3.36×10²² | 4.37×10²² | 1.56×10³³ | 1.76×10³³ | 2.03×10³² | 1.28×10⁻¹ | 1.08×10⁻¹ |
| | | | 3.83 | 102.86 | 3.03 | 66.83 | 92.83 | 1.22×10²² | 1.04×10²² | 7.19×10³² | 8.27×10³² | 1.58×10³² | 9.74×10⁻² | 7.79×10⁻² |
| Dro | 108 / 86 | 2.04 / 2.06 | 1.00 | 10.00 | 2.00 | 1220.60 | 479.55 | 1.61×10²¹ | 3.47×10²¹ | 2.89×10³¹ | 5.69×10³¹ | 4.74×10³⁰ | 1.04×10⁻² | 1.03×10⁻² |
| | | | 13.00 | 90.00 | 10.00 | 2607.50 | 2305.85 | 2.25×10²² | 3.70×10²² | 7.86×10³² | 8.07×10³² | 3.21×10³² | 1.55×10⁰ | 6.08×10⁻¹ |
| | | | 5.91 | 31.25 | 5.23 | 1856.77 | 1030.71 | 8.22×10²¹ | 1.52×10²² | 2.16×10³² | 2.71×10³² | 5.50×10³¹ | 3.96×10⁻¹ | 2.54×10⁻¹ |
| | | | 2.61 | 12.92 | 1.70 | 289.59 | 258.61 | 5.05×10²¹ | 7.40×10²¹ | 1.62×10³² | 1.67×10³² | 3.86×10³¹ | 3.01×10⁻¹ | 1.40×10⁻¹ |

| | | | | | | | | | | | | | |
|---|---|---|---|---|---|---|---|---|---|---|---|---|---|
| Dri | 22 / 10 | 0.42 / 0.24 | 3.00 24.00 11.59 5.55 | 20.00 240.00 57.73 53.80 | 3.00 10.00 5.91 1.93 | 1562.14 2926.13 2158.72 341.51 | 753.18 2823.26 1296.86 428.60 | $5.13 \times 10^{21}$ $3.50 \times 10^{22}$ $1.61 \times 10^{22}$ $8.77 \times 10^{21}$ | $1.21 \times 10^{22}$ $6.41 \times 10^{22}$ $2.82 \times 10^{22}$ $1.36 \times 10^{22}$ | $8.99 \times 10^{31}$ $1.25 \times 10^{33}$ $4.80 \times 10^{32}$ $3.17 \times 10^{32}$ | $1.30 \times 10^{32}$ $1.28 \times 10^{33}$ $5.57 \times 10^{32}$ $3.20 \times 10^{32}$ | $6.50 \times 10^{30}$ $2.97 \times 10^{32}$ $7.67 \times 10^{31}$ $7.75 \times 10^{31}$ | $1.44 \times 10^{-2}$ $1.01 \times 10^{0}$ $2.23 \times 10^{-1}$ $2.27 \times 10^{-1}$ | $1.42 \times 10^{-2}$ $5.04 \times 10^{-1}$ $1.61 \times 10^{-1}$ $1.25 \times 10^{-1}$ |
| Dso | 289 / 237 | 5.47 / 5.67 | 2.00 22.00 7.26 3.61 | 10.00 260.00 94.43 54.83 | 2.00 11.00 6.69 2.06 | 622.39 3075.18 2180.54 339.50 | 291.43 1910.75 1179.57 218.98 | $6.96 \times 10^{20}$ $7.44 \times 10^{22}$ $1.35 \times 10^{22}$ $8.47 \times 10^{21}$ | $8.81 \times 10^{20}$ $6.45 \times 10^{22}$ $2.22 \times 10^{22}$ $1.11 \times 10^{22}$ | $1.92 \times 10^{31}$ $2.80 \times 10^{33}$ $4.28 \times 10^{32}$ $3.20 \times 10^{32}$ | $6.35 \times 10^{31}$ $3.53 \times 10^{33}$ $5.24 \times 10^{32}$ $3.83 \times 10^{32}$ | $1.89 \times 10^{29}$ $5.58 \times 10^{32}$ $9.55 \times 10^{31}$ $1.10 \times 10^{32}$ | $3.65 \times 10^{-4}$ $2.43 \times 10^{0}$ $2.10 \times 10^{-1}$ $3.59 \times 10^{-1}$ | $3.65 \times 10^{-4}$ $7.08 \times 10^{-1}$ $1.97 \times 10^{-1}$ $1.52 \times 10^{-1}$ |
| Dsi | 118 / 96 | 2.23 / 2.30 | 2.00 29.00 13.16 4.60 | 30.00 330.00 140.59 57.02 | 4.00 14.00 7.69 1.78 | 1666.83 3010.57 2319.57 285.93 | 847.80 2149.49 1295.03 204.27 | $3.24 \times 10^{21}$ $5.09 \times 10^{22}$ $1.69 \times 10^{22}$ $7.86 \times 10^{21}$ | $7.29 \times 10^{21}$ $7.95 \times 10^{22}$ $3.03 \times 10^{22}$ $1.11 \times 10^{22}$ | $6.48 \times 10^{31}$ $2.13 \times 10^{33}$ $5.81 \times 10^{32}$ $3.24 \times 10^{32}$ | $1.20 \times 10^{32}$ $2.65 \times 10^{33}$ $7.18 \times 10^{32}$ $4.07 \times 10^{32}$ | $4.97 \times 10^{28}$ $8.99 \times 10^{32}$ $1.37 \times 10^{32}$ $1.56 \times 10^{32}$ | $1.09 \times 10^{-4}$ $1.13 \times 10^{0}$ $2.56 \times 10^{-1}$ $2.45 \times 10^{-1}$ | $1.09 \times 10^{-4}$ $5.30 \times 10^{-1}$ $1.79 \times 10^{-1}$ $1.32 \times 10^{-1}$ |
| Dsc | 10 / 6 | 0.19 / 0.14 | 5.00 24.00 14.60 6.04 | 50.00 280.00 164.00 65.52 | 4.00 10.00 7.70 1.83 | 1996.60 3089.01 2521.67 279.58 | 1036.19 1813.91 1452.31 253.95 | $8.86 \times 10^{21}$ $2.99 \times 10^{22}$ $2.12 \times 10^{22}$ $6.62 \times 10^{21}$ | $2.04 \times 10^{22}$ $7.29 \times 10^{22}$ $4.72 \times 10^{22}$ $1.72 \times 10^{22}$ | $2.65 \times 10^{32}$ $1.46 \times 10^{33}$ $7.51 \times 10^{32}$ $3.54 \times 10^{32}$ | $2.80 \times 10^{32}$ $1.71 \times 10^{33}$ $9.73 \times 10^{32}$ $4.69 \times 10^{32}$ | $1.54 \times 10^{31}$ $8.15 \times 10^{32}$ $2.22 \times 10^{32}$ $2.47 \times 10^{32}$ | $3.59 \times 10^{-2}$ $9.10 \times 10^{-1}$ $2.98 \times 10^{-1}$ $2.91 \times 10^{-1}$ | $3.47 \times 10^{-2}$ $4.76 \times 10^{-1}$ $2.00 \times 10^{-1}$ $1.51 \times 10^{-1}$ |
| Dao | 422 / 349 | 7.98 / 8.35 | 2.00 34.00 8.36 3.68 | 10.00 320.00 96.69 55.09 | 2.00 11.00 6.70 1.99 | 1404.19 3211.27 2147.11 280.40 | 643.66 2433.42 1193.44 230.28 | $2.13 \times 10^{21}$ $5.71 \times 10^{22}$ $1.38 \times 10^{22}$ $8.26 \times 10^{21}$ | $5.12 \times 10^{21}$ $7.77 \times 10^{22}$ $2.33 \times 10^{22}$ $1.11 \times 10^{22}$ | $4.42 \times 10^{31}$ $2.62 \times 10^{33}$ $4.37 \times 10^{32}$ $3.30 \times 10^{32}$ | $8.26 \times 10^{31}$ $2.69 \times 10^{33}$ $5.28 \times 10^{32}$ $3.73 \times 10^{32}$ | $3.56 \times 10^{29}$ $7.01 \times 10^{32}$ $9.05 \times 10^{31}$ $1.01 \times 10^{32}$ | $1.01 \times 10^{-2}$ $1.51 \times 10^{0}$ $2.57 \times 10^{-1}$ $2.39 \times 10^{-1}$ | $1.01 \times 10^{-3}$ $6.01 \times 10^{-1}$ $1.81 \times 10^{-1}$ $1.26 \times 10^{-1}$ |
| Dai | 379 / 309 | 7.17 / 7.39 | 2.00 40.00 13.07 4.96 | 20.00 400.00 131.29 61.55 | 3.00 13.00 7.62 1.75 | 1325.89 3016.22 2217.79 288.39 | 670.47 2971.44 1314.03 282.56 | $4.47 \times 10^{21}$ $6.74 \times 10^{22}$ $1.72 \times 10^{22}$ $8.49 \times 10^{21}$ | $8.06 \times 10^{21}$ $1.29 \times 10^{22}$ $3.25 \times 10^{22}$ $1.41 \times 10^{22}$ | $7.71 \times 10^{31}$ $2.71 \times 10^{33}$ $5.80 \times 10^{32}$ $3.42 \times 10^{32}$ | $1.51 \times 10^{32}$ $3.42 \times 10^{33}$ $7.27 \times 10^{32}$ $4.36 \times 10^{32}$ | $1.16 \times 10^{29}$ $9.79 \times 10^{32}$ $1.47 \times 10^{32}$ $1.55 \times 10^{32}$ | $2.60 \times 10^{-4}$ $1.82 \times 10^{0}$ $2.72 \times 10^{-1}$ $2.68 \times 10^{-1}$ | $2.60 \times 10^{-4}$ $6.45 \times 10^{-1}$ $1.87 \times 10^{-1}$ $1.34 \times 10^{-1}$ |
| Dac | 82 / 62 | 1.55 / 1.48 | 5.00 32.00 14.94 6.46 | 30.00 280.00 156.59 56.18 | 4.00 10.00 7.73 1.69 | 1751.71 3040.00 2260.35 249.51 | 918.88 3267.50 1425.73 380.10 | $6.83 \times 10^{21}$ $1.14 \times 10^{23}$ $2.37 \times 10^{22}$ $1.55 \times 10^{22}$ | $1.19 \times 10^{22}$ $8.53 \times 10^{22}$ $4.08 \times 10^{22}$ $1.68 \times 10^{22}$ | $1.72 \times 10^{32}$ $5.80 \times 10^{33}$ $8.51 \times 10^{32}$ $7.64 \times 10^{32}$ | $2.23 \times 10^{32}$ $5.98 \times 10^{33}$ $1.07 \times 10^{33}$ $8.58 \times 10^{32}$ | $3.84 \times 10^{30}$ $1.08 \times 10^{33}$ $2.18 \times 10^{32}$ $2.37 \times 10^{32}$ | $5.06 \times 10^{-3}$ $1.56 \times 10^{0}$ $2.84 \times 10^{-1}$ $2.89 \times 10^{-1}$ | $5.04 \times 10^{-3}$ $6.10 \times 10^{-1}$ $1.91 \times 10^{-1}$ $1.42 \times 10^{-1}$ |
| Dho | 23 / 12 | 0.43 / 0.29 | 3.00 23.00 10.04 5.43 | 190.00 500.00 285.22 64.94 | 5.00 11.00 8.91 1.38 | 2007.66 3280.09 2687.12 310.07 | 1023.41 1767.06 1462.18 200.45 | $1.03 \times 10^{22}$ $4.19 \times 10^{22}$ $2.46 \times 10^{22}$ $1.00 \times 10^{22}$ | $1.60 \times 10^{22}$ $5.80 \times 10^{22}$ $3.49 \times 10^{22}$ $1.19 \times 10^{22}$ | $3.42 \times 10^{32}$ $2.09 \times 10^{33}$ $1.08 \times 10^{33}$ $4.98 \times 10^{32}$ | $3.83 \times 10^{32}$ $2.28 \times 10^{33}$ $1.27 \times 10^{33}$ $5.45 \times 10^{32}$ | $1.12 \times 10^{31}$ $4.05 \times 10^{32}$ $1.82 \times 10^{32}$ $1.22 \times 10^{32}$ | $1.35 \times 10^{-2}$ $4.66 \times 10^{-1}$ $1.78 \times 10^{-1}$ $1.27 \times 10^{-1}$ | $1.34 \times 10^{-2}$ $3.18 \times 10^{-1}$ $1.42 \times 10^{-1}$ $8.43 \times 10^{-2}$ |
| Dhi | 21 / 10 | 0.40 / 0.24 | 7.00 27.00 15.29 5.93 | 120.00 490.00 329.05 83.84 | 5.00 10.00 8.81 1.40 | 2182.36 3298.74 2703.99 259.93 | 1175.45 2729.12 1578.22 355.64 | $9.29 \times 10^{21}$ $8.58 \times 10^{22}$ $2.93 \times 10^{22}$ $1.61 \times 10^{22}$ | $2.31 \times 10^{22}$ $1.01 \times 10^{22}$ $4.58 \times 10^{22}$ $1.74 \times 10^{22}$ | $3.13 \times 10^{32}$ $5.11 \times 10^{33}$ $1.28 \times 10^{33}$ $9.87 \times 10^{32}$ | $4.11 \times 10^{32}$ $5.52 \times 10^{33}$ $1.50 \times 10^{33}$ $1.05 \times 10^{33}$ | $1.18 \times 10^{31}$ $6.52 \times 10^{32}$ $2.20 \times 10^{32}$ $1.53 \times 10^{32}$ | $1.07 \times 10^{-2}$ $5.06 \times 10^{-1}$ $2.13 \times 10^{-1}$ $1.52 \times 10^{-1}$ | $1.06 \times 10^{-2}$ $3.36 \times 10^{-1}$ $1.64 \times 10^{-1}$ $9.85 \times 10^{-2}$ |
| Dhc | 11 / 8 | 0.21 / 0.19 | 8.00 40.00 18.18 11.11 | 260.00 510.00 310.91 70.92 | 5.00 10.00 8.82 1.47 | 2207.91 2975.93 2619.35 225.29 | 1186.99 2277.00 1679.71 370.00 | $1.41 \times 10^{22}$ $9.09 \times 10^{22}$ $3.40 \times 10^{22}$ $2.11 \times 10^{22}$ | $2.73 \times 10^{22}$ $9.62 \times 10^{22}$ $5.11 \times 10^{22}$ $2.00 \times 10^{22}$ | $6.02 \times 10^{32}$ $3.46 \times 10^{33}$ $1.34 \times 10^{33}$ $7.89 \times 10^{32}$ | $6.81 \times 10^{32}$ $4.03 \times 10^{33}$ $1.73 \times 10^{33}$ $9.83 \times 10^{32}$ | $6.89 \times 10^{31}$ $1.02 \times 10^{33}$ $3.91 \times 10^{32}$ $3.16 \times 10^{32}$ | $6.47 \times 10^{-2}$ $6.90 \times 10^{-1}$ $2.87 \times 10^{-1}$ $1.95 \times 10^{-1}$ | $6.08 \times 10^{-2}$ $4.08 \times 10^{-1}$ $2.08 \times 10^{-1}$ $1.08 \times 10^{-1}$ |
| Dko | 38 / 23 | 0.72 / 0.55 | 4.00 20.00 10.76 4.34 | 250.00 600.00 328.95 75.54 | 4.00 12.00 7.68 1.96 | 1991.04 3660.87 2670.97 353.10 | 1095.86 2825.28 1623.08 413.42 | $1.21 \times 10^{22}$ $6.68 \times 10^{22}$ $2.81 \times 10^{22}$ $1.36 \times 10^{22}$ | $1.18 \times 10^{22}$ $7.58 \times 10^{22}$ $3.93 \times 10^{22}$ $1.35 \times 10^{22}$ | $3.64 \times 10^{32}$ $3.89 \times 10^{33}$ $1.21 \times 10^{33}$ $7.26 \times 10^{32}$ | $4.84 \times 10^{32}$ $4.00 \times 10^{33}$ $1.34 \times 10^{33}$ $7.82 \times 10^{32}$ | $7.77 \times 10^{29}$ $6.86 \times 10^{32}$ $1.30 \times 10^{32}$ $1.36 \times 10^{32}$ | $9.15 \times 10^{-4}$ $4.39 \times 10^{-1}$ $1.25 \times 10^{-1}$ $1.14 \times 10^{-1}$ | $9.14 \times 10^{-4}$ $3.05 \times 10^{-1}$ $1.03 \times 10^{-1}$ $8.36 \times 10^{-2}$ |
| Dki | 82 / 62 | 1.55 / 1.48 | 7.00 37.00 15.85 | 230.00 600.00 334.15 | 5.00 10.00 8.70 | 1996.44 3468.65 2544.71 | 1098.96 2482.48 1538.41 | $1.02 \times 10^{22}$ $5.97 \times 10^{22}$ $2.78 \times 10^{22}$ | $1.93 \times 10^{22}$ $8.36 \times 10^{22}$ $4.71 \times 10^{22}$ | $3.42 \times 10^{32}$ $2.56 \times 10^{33}$ $1.15 \times 10^{33}$ | $4.65 \times 10^{32}$ $3.51 \times 10^{33}$ $1.38 \times 10^{33}$ | $9.72 \times 10^{29}$ $9.42 \times 10^{32}$ $2.39 \times 10^{32}$ | $6.24 \times 10^{-4}$ $1.09 \times 10^{0}$ $2.26 \times 10^{-1}$ | $6.23 \times 10^{-4}$ $5.21 \times 10^{-1}$ $1.65 \times 10^{-1}$ |

| | | | | | | | | | | | | | |
|---|---|---|---|---|---|---|---|---|---|---|---|---|---|
| | | | 5.45 | 73.94 | 1.31 | 289.60 | 328.69 | 9.96×10²¹ | 1.38×10²² | 4.67×10³² | 5.66×10³² | 2.08×10³² | 2.15×10⁻¹ | 1.16×10⁻¹ |
| Dkc | 85 / 65 | 1.61 / 1.55 | 3.0 | 160.00 | 3.00 | 2015.14 | 1106.58 | 1.18×10²² | 2.02×10²² | 3.54×10³² | 4.72×10³² | 8.44×10³⁰ | 3.92×10⁻³ | 3.91×10⁻³ |
| | | | 57.00 | 1270.00 | 10.00 | 3509.50 | 3579.25 | 1.01×10²³ | 1.05×10²² | 3.82×10³³ | 5.46×10³³ | 2.86×10³³ | 1.63×10⁰ | 6.19×10⁻¹ |
| | | | 20.16 | 431.53 | 8.31 | 2649.96 | 1750.22 | 3.32×10²² | 5.72×10²² | 1.42×10³³ | 1.80×10³³ | 3.76×10³² | 2.50×10⁻¹ | 1.75×10⁻¹ |
| | | | 9.19 | 199.00 | 1.73 | 331.94 | 532.53 | 1.29×10²² | 1.93×10²² | 6.39×10³² | 1.00×10³³ | 5.00×10³² | 2.62×10⁻¹ | 1.26×10⁻¹ |
| Ero | 5 / 4 | 0.09 / 0.10 | 5.00 | 30.00 | 12.00 | 1008.40 | 485.16 | 8.69×10²¹ | 1.94×10²² | 2.13×10³² | 4.46×10³² | 2.24×10³² | 1.73×10⁻¹ | 1.47×10⁻¹ |
| | | | 15.00 | 120.00 | 14.00 | 2234.97 | 1208.88 | 5.00×10²² | 3.47×10²² | 1.91×10³³ | 2.56×10³³ | 6.62×10³² | 1.09×10⁰ | 5.22×10⁻¹ |
| | | | 10.80 | 60.00 | 13.00 | 1725.60 | 914.52 | 3.05×10²² | 2.77×10²² | 1.15×10³³ | 1.58×10³³ | 4.30×10³² | 5.03×10⁻¹ | 3.10×10⁻¹ |
| | | | 3.70 | 37.42 | 0.71 | 472.25 | 293.51 | 1.54×10²² | 5.79×10²¹ | 6.59×10³² | 8.32×10³² | 2.15×10³² | 3.50×10⁻¹ | 1.37×10⁻¹ |
| Eri | 1 / 1 | 0.02 / 0.02 | 15.00 | 80.00 | 11.00 | 2417.44 | 1518.98 | 4.44×10²² | 7.52×10²² | 1.87×10³³ | 2.04×10³³ | 1.67×10³² | 8.91×10⁻² | 8.18×10⁻² |
| | | | 15.00 | 80.00 | 11.00 | 2417.44 | 1518.98 | 4.44×10²² | 7.52×10²² | 1.87×10³³ | 2.04×10³³ | 1.67×10³² | 8.91×10⁻² | 8.18×10⁻² |
| | | | 15.00 | 80.00 | 11.00 | 2417.44 | 1518.98 | 4.44×10²² | 7.52×10²² | 1.87×10³³ | 2.04×10³³ | 1.67×10³² | 8.91×10⁻² | 8.18×10⁻² |
| | | | - | - | - | - | - | - | - | - | - | - | - | - |
| Eso | 58 / 40 | 1.10 / 0.96 | 3.00 | 30.00 | 9.00 | 1450.76 | 616.84 | 5.00×10²¹ | 1.08×10²² | 1.33×10³² | 2.39×10³² | 4.05×10³⁰ | 1.35×10⁻² | 1.34×10⁻² |
| | | | 22.00 | 240.00 | 15.00 | 2919.11 | 1642.79 | 5.73×10²² | 5.50×10²² | 2.40×10³³ | 2.79×10³³ | 7.28×10³² | 1.15×10⁰ | 5.35×10⁻¹ |
| | | | 10.28 | 153.28 | 12.07 | 2352.86 | 1253.99 | 2.18×10²² | 2.76×10²² | 8.59×10³² | 1.13×10³³ | 2.70×10³² | 3.88×10⁻¹ | 2.54×10⁻¹ |
| | | | 5.27 | 54.85 | 1.24 | 348.79 | 204.82 | 1.05×10²² | 1.02×10²² | 4.96×10³² | 5.50×10³² | 1.64×10³² | 2.68×10⁻¹ | 1.38×10⁻¹ |
| Esi | 64 / 44 | 1.21 / 1.05 | 7.00 | 70.00 | 7.00 | 1450.21 | 820.49 | 9.28×10²¹ | 1.16×10²² | 2.47×10³² | 3.70×10³² | 3.55×10³⁰ | 3.63×10⁻³ | 3.61×10⁻³ |
| | | | 35.00 | 360.00 | 15.00 | 2980.07 | 1789.9 | 5.53×10²² | 9.22×10²² | 3.04×10³³ | 3.43×10³³ | 1.11×10³³ | 9.07×10⁻¹ | 4.76×10⁻¹ |
| | | | 17.11 | 177.81 | 12.19 | 2420.42 | 1335.52 | 2.85×10²² | 3.94×10²² | 1.15×10³³ | 1.44×10³³ | 2.89×10³² | 3.26×10⁻¹ | 2.21×10⁻¹ |
| | | | 6.58 | 58.51 | 1.33 | 347.74 | 190.48 | 1.14×10²² | 1.49×10²² | 6.14×10³² | 6.58×10³² | 1.99×10³² | 2.53×10⁻¹ | 1.36×10⁻¹ |
| Esc | 0 | 0 | - | - | - | - | - | - | - | - | - | - | - | - |
| Eao | 43 / 28 | 0.81 / 0.67 | 2.00 | 50.00 | 11.00 | 1467.74 | 757.18 | 9.72×10²¹ | 1.35×10²² | 2.29×10³² | 4.87×10³² | 2.10×10³¹ | 2.88×10⁻² | 2.80×10⁻² |
| | | | 24.00 | 300.00 | 15.00 | 3066.65 | 1757.29 | 6.00×10²² | 6.79×10²² | 2.89×10³³ | 3.65×10³³ | 7.53×10³² | 1.12×10⁰ | 5.29×10⁻¹ |
| | | | 10.44 | 148.14 | 11.86 | 2172.16 | 1207.94 | 2.92×10²² | 3.36×10²² | 8.72×10³² | 1.12×10³³ | 2.51×10³² | 3.77×10⁻¹ | 2.51×10⁻¹ |
| | | | 4.56 | 60.92 | 1.01 | 341.76 | 217.02 | 1.89×10²² | 1.05×10²² | 6.00×10³² | 6.76×10³² | 1.44×10³² | 2.59×10⁻¹ | 1.29×10⁻¹ |
| Eai | 122 / 100 | 2.31 / 2.39 | 8.00 | 60.00 | 11.00 | 1316.63 | 571.62 | 9.83×10²¹ | 1.66×10²² | 2.34×10³² | 4.24×10³² | 7.66×10³⁰ | 1.71×10⁻² | 1.68×10⁻² |
| | | | 39.00 | 470.00 | 15.00 | 2935.28 | 2432.55 | 7.43×10²² | 7.89×10²² | 3.31×10³³ | 3.78×10³³ | 7.53×10³² | 1.23×10⁰ | 5.51×10⁻¹ |
| | | | 18.52 | 179.59 | 12.11 | 2242.33 | 1297.46 | 2.94×10²² | 4.26×10²² | 1.15×10³³ | 1.44×10³³ | 2.84×10³² | 3.16×10⁻¹ | 2.21×10⁻¹ |
| | | | 6.34 | 64.75 | 1.11 | 318.13 | 268.88 | 1.35×10²² | 1.49×10²² | 6.53×10³² | 6.87×10³² | 1.53×10³² | 2.29×10⁻¹ | 1.16×10⁻¹ |
| Eac | 64 / 44 | 1.21 / 1.05 | 10.00 | 60.00 | 10.00 | 1264.03 | 851.94 | 1.06×10²² | 1.35×10²² | 2.66×10³² | 4.32×10³² | 3.35×10³⁰ | 2.78×10⁻³ | 2.78×10⁻³ |
| | | | 58.00 | 620.00 | 15.00 | 3058.38 | 2570.65 | 5.59×10²² | 9.60×10²² | 2.68×10³³ | 2.88×10³³ | 9.07×10³² | 9.51×10⁻¹ | 4.87×10⁻¹ |
| | | | 23.25 | 211.09 | 12.06 | 2289.81 | 1466.52 | 2.95×10²² | 5.00×10²² | 1.16×10³³ | 1.50×10³³ | 3.39×10³² | 3.39×10⁻¹ | 2.39×10⁻¹ |
| | | | 8.24 | 78.12 | 1.22 | 326.21 | 387.21 | 1.05×10²² | 1.55×10²² | 5.21×10³² | 5.75×10³² | 1.67×10³² | 1.92×10⁻¹ | 1.03×10⁻¹ |
| Eho | 32 / 17 | 0.61 / 0.41 | 4.00 | 210.00 | 10.00 | 1784.92 | 1033.40 | 1.87×10²² | 1.67×10²² | 8.65×10³² | 1.04×10³³ | 3.25×10³⁰ | 1.63×10⁻³ | 1.63×10⁻³ |
| | | | 22.00 | 530.00 | 15.00 | 3158.91 | 2127.30 | 7.60×10²² | 7.90×10²² | 3.68×10³³ | 4.19×10³³ | 6.01×10³² | 4.35×10⁻¹ | 3.03×10⁻¹ |
| | | | 10.19 | 348.13 | 12.75 | 2683.90 | 1499.07 | 3.26×10²² | 3.72×10²² | 1.71×10³³ | 1.94×10³³ | 2.29×10³² | 1.49×10⁻¹ | 1.23×10⁻¹ |
| | | | 4.39 | 89.06 | 1.24 | 297.05 | 195.50 | 1.16×10²² | 1.49×10²² | 5.78×10³² | 6.05×10³² | 1.35×10³² | 1.02×10⁻¹ | 7.31×10⁻² |
| Ehi | 41 / 26 | 0.78 / 0.62 | 8.00 | 250.00 | 10.00 | 2279.54 | 1183.13 | 1.90×10²² | 1.84×10²² | 6.26×10³² | 9.42×10³² | 3.13×10³⁰ | 2.04×10⁻³ | 2.03×10⁻³ |
| | | | 31.00 | 880.00 | 15.00 | 3427.30 | 3005.58 | 1.06×10²³ | 1.01×10²³ | 6.33×10³³ | 7.32×10³³ | 9.83×10³² | 9.61×10⁻¹ | 4.90×10⁻¹ |
| | | | 17.76 | 368.90 | 12.29 | 2691.93 | 1552.72 | 3.63×10²² | 4.74×10²² | 1.78×10³³ | 2.06×10³³ | 2.85×10³² | 1.95×10⁻¹ | 1.50×10⁻¹ |
| | | | 5.14 | 131.02 | 1.33 | 302.18 | 338.52 | 1.61×10²² | 1.69×10²² | 1.12×10³³ | 1.20×10³³ | 2.04×10³² | 1.72×10⁻¹ | 1.00×10⁻¹ |
| Ehc | 19 / 10 | 0.36 / 0.24 | 12.00 | 270.00 | 11.00 | 2034.35 | 1186.33 | 2.12×10²² | 3.11×10²² | 1.03×10³³ | 1.18×10³³ | 5.03×10³¹ | 3.23×10⁻² | 3.13×10⁻² |
| | | | 59.00 | 920.00 | 15.00 | 3462.71 | 2462.96 | 5.38×10²² | 7.88×10²² | 2.85×10³³ | 3.78×10³³ | 9.50×10³² | 4.75×10⁻¹ | 3.22×10⁻¹ |
| | | | 25.21 | 458.95 | 12.26 | 2819.45 | 1683.01 | 3.96×10²² | 5.69×10²² | 1.94×10³³ | 2.28×10³³ | 3.40×10³² | 1.77×10⁻¹ | 1.40×10⁻¹ |
| | | | 13.12 | 197.42 | 1.41 | 350.85 | 296.07 | 8.04×10²² | 1.20×10²² | 4.90×10³² | 6.12×10³² | 2.75×10³² | 1.40×10⁻¹ | 9.37×10⁻² |

| | | | | | | | | | | | | | |
|---|---|---|---|---|---|---|---|---|---|---|---|---|---|
| Eko | 47 / 32 | 0.89 / 0.77 | 4.00 28.00 12.47 4.86 | 250.00 930.00 427.45 155.34 | 11.00 15.00 12.23 1.29 | 2192.65 3119.90 2676.24 238.48 | 1136.90 2691.13 1489.64 256.57 | $1.70 \times 10^{22}$ $1.13 \times 10^{23}$ $3.48 \times 10^{22}$ $1.62 \times 10^{22}$ | $2.31 \times 10^{22}$ $8.48 \times 10^{22}$ $4.45 \times 10^{22}$ $1.55 \times 10^{22}$ | $6.53 \times 10^{32}$ $7.50 \times 10^{33}$ $1.85 \times 10^{33}$ $1.17 \times 10^{33}$ | $8.63 \times 10^{32}$ $7.92 \times 10^{33}$ $2.10 \times 10^{33}$ $1.18 \times 10^{33}$ | $6.34 \times 10^{31}$ $6.66 \times 10^{32}$ $2.50 \times 10^{32}$ $1.38 \times 10^{32}$ | $3.21 \times 10^{-2}$ $5.39 \times 10^{-1}$ $1.74 \times 10^{-1}$ $1.27 \times 10^{-1}$ | $3.11 \times 10^{-2}$ $3.50 \times 10^{-1}$ $1.40 \times 10^{-1}$ $8.54 \times 10^{-2}$ |
| Eki | 109 / 87 | 2.06 / 2.08 | 8.00 51.00 20.66 6.90 | 160.00 930.00 419.17 164.10 | 10.00 15.00 12.54 1.36 | 1793.88 3345.05 2656.90 319.18 | 1113.22 2683.44 1553.89 260.09 | $1.77 \times 10^{22}$ $8.92 \times 10^{22}$ $3.77 \times 10^{22}$ $1.58 \times 10^{22}$ | $2.01 \times 10^{22}$ $1.07 \times 10^{23}$ $5.37 \times 10^{22}$ $1.77 \times 10^{22}$ | $6.75 \times 10^{32}$ $4.85 \times 10^{33}$ $1.92 \times 10^{33}$ $1.00 \times 10^{33}$ | $8.19 \times 10^{32}$ $5.28 \times 10^{33}$ $2.25 \times 10^{33}$ $1.04 \times 10^{32}$ | $4.32 \times 10^{30}$ $1.05 \times 10^{33}$ $3.31 \times 10^{32}$ $2.13 \times 10^{32}$ | $4.87 \times 10^{-3}$ $8.62 \times 10^{-1}$ $2.12 \times 10^{-1}$ $1.71 \times 10^{-1}$ | $4.84 \times 10^{-3}$ $4.63 \times 10^{-1}$ $1.61 \times 10^{-1}$ $9.93 \times 10^{-2}$ |
| Ekc | 250 / 207 | 4.73 / 4.95 | 7.00 84.00 26.62 11.41 | 220.00 1250.00 506.48 207.58 | 8.00 16.00 12.72 1.39 | 1690.34 3488.98 2599.49 337.03 | 1008.24 4166.29 1624.92 388.42 | $1.90 \times 10^{22}$ $1.04 \times 10^{23}$ $4.33 \times 10^{22}$ $1.48 \times 10^{22}$ | $2.84 \times 10^{22}$ $1.07 \times 10^{23}$ $6.30 \times 10^{22}$ $1.62 \times 10^{22}$ | $6.53 \times 10^{32}$ $6.00 \times 10^{33}$ $2.22 \times 10^{33}$ $9.46 \times 10^{32}$ | $9.53 \times 10^{32}$ $6.21 \times 10^{33}$ $2.64 \times 10^{33}$ $9.98 \times 10^{32}$ | $7.07 \times 10^{30}$ $2.22 \times 10^{33}$ $4.23 \times 10^{32}$ $2.99 \times 10^{32}$ | $6.84 \times 10^{-3}$ $1.14 \times 10^{0}$ $2.23 \times 10^{-1}$ $1.76 \times 10^{-1}$ | $6.79 \times 10^{-3}$ $5.33 \times 10^{-1}$ $1.68 \times 10^{-1}$ $1.01 \times 10^{-1}$ |
| Fro | 0 | 0 | - | - | - | - | - | - | - | - | - | - | - | - |
| Fri | 0 | 0 | - | - | - | - | - | - | - | - | - | - | - | - |
| Fso | 9 / 6 | 0.17 / 0.14 | 9.00 18.00 14.67 3.16 | 60.00 240.00 142.22 73.45 | 16.00 19.00 17.33 1.32 | 1764.76 2681.05 2215.58 367.48 | 915.07 1380.16 1165.72 169.98 | $1.65 \times 10^{22}$ $6.22 \times 10^{22}$ $3.14 \times 10^{22}$ $1.66 \times 10^{22}$ | $1.86 \times 10^{22}$ $7.31 \times 10^{22}$ $3.07 \times 10^{22}$ $1.66 \times 10^{22}$ | $7.14 \times 10^{32}$ $3.29 \times 10^{33}$ $1.46 \times 10^{33}$ $8.04 \times 10^{32}$ | $1.04 \times 10^{33}$ $3.31 \times 10^{33}$ $1.82 \times 10^{33}$ $7.95 \times 10^{32}$ | $2.43 \times 10^{31}$ $8.80 \times 10^{32}$ $3.51 \times 10^{32}$ $2.46 \times 10^{32}$ | $7.39 \times 10^{-3}$ $5.27 \times 10^{-1}$ $2.98 \times 10^{-1}$ $1.81 \times 10^{-1}$ | $7.33 \times 10^{-3}$ $3.45 \times 10^{-1}$ $2.15 \times 10^{-1}$ $1.17 \times 10^{-1}$ |
| Fsi | 4 / 2 | 0.08 / 0.05 | 16.00 25.00 21.00 4.69 | 80.00 260.00 202.50 83.42 | 16.00 22.00 18.00 2.71 | 1992.75 2634.51 2390.46 277.26 | 1132.17 1770.57 1375.85 274.83 | $2.28 \times 10^{22}$ $4.71 \times 10^{22}$ $3.05 \times 10^{22}$ $1.13 \times 10^{22}$ | $2.92 \times 10^{22}$ $5.69 \times 10^{22}$ $4.23 \times 10^{22}$ $1.30 \times 10^{22}$ | $9.71 \times 10^{32}$ $2.57 \times 10^{33}$ $1.47 \times 10^{33}$ $7.46 \times 10^{32}$ | $1.26 \times 10^{33}$ $2.93 \times 10^{33}$ $2.13 \times 10^{33}$ $7.74 \times 10^{32}$ | $1.80 \times 10^{32}$ $1.36 \times 10^{33}$ $6.63 \times 10^{32}$ $5.22 \times 10^{32}$ | $1.39 \times 10^{-1}$ $1.08 \times 10^{0}$ $5.42 \times 10^{-1}$ $4.67 \times 10^{-1}$ | $1.22 \times 10^{-1}$ $5.20 \times 10^{-1}$ $3.06 \times 10^{-1}$ $2.03 \times 10^{-1}$ |
| Fsc | 3 / 2 | 0.06 / 0.05 | 21.00 42.00 34.33 11.59 | 220.00 320.00 260.00 52.92 | 16.00 16.00 16.00 0.00 | 2108.51 2558.42 2367.28 232.45 | 1321.26 1521.06 1431.19 101.39 | $3.12 \times 10^{22}$ $3.87 \times 10^{22}$ $3.42 \times 10^{22}$ $3.99 \times 10^{21}$ | $3.06 \times 10^{22}$ $8.48 \times 10^{22}$ $5.70 \times 10^{22}$ $2.71 \times 10^{22}$ | $1.09 \times 10^{33}$ $2.17 \times 10^{33}$ $1.58 \times 10^{33}$ $5.46 \times 10^{32}$ | $1.80 \times 10^{33}$ $2.41 \times 10^{33}$ $2.08 \times 10^{33}$ $3.06 \times 10^{32}$ | $2.40 \times 10^{32}$ $7.11 \times 10^{32}$ $5.03 \times 10^{32}$ $2.41 \times 10^{32}$ | $1.11 \times 10^{-1}$ $6.52 \times 10^{-1}$ $3.80 \times 10^{-1}$ $2.71 \times 10^{-1}$ | $9.95 \times 10^{-2}$ $3.95 \times 10^{-1}$ $2.56 \times 10^{-1}$ $1.48 \times 10^{-1}$ |
| Fao | 5 / 4 | 0.09 / 0.10 | 7.00 23.00 13.80 6.06 | 80.00 210.00 120.00 51.96 | 16.00 17.00 16.40 0.55 | 1851.38 2444.01 2154.02 238.36 | 1011.76 1545.93 1228.76 196.19 | $1.22 \times 10^{22}$ $3.67 \times 10^{22}$ $2.57 \times 10^{22}$ $8.84 \times 10^{21}$ | $2.30 \times 10^{22}$ $3.81 \times 10^{22}$ $3.07 \times 10^{22}$ $7.11 \times 10^{22}$ | $3.26 \times 10^{32}$ $1.99 \times 10^{33}$ $1.19 \times 10^{33}$ $6.16 \times 10^{32}$ | $7.81 \times 10^{32}$ $2.03 \times 10^{33}$ $1.45 \times 10^{33}$ $4.88 \times 10^{32}$ | $4.13 \times 10^{31}$ $4.55 \times 10^{32}$ $2.62 \times 10^{31}$ $1.62 \times 10^{32}$ | $2.07 \times 10^{-2}$ $1.40 \times 10^{0}$ $4.20 \times 10^{-1}$ $5.55 \times 10^{-1}$ | $2.03 \times 10^{-2}$ $5.82 \times 10^{-1}$ $2.30 \times 10^{-1}$ $2.12 \times 10^{-1}$ |
| Fai | 8 / 4 | 0.15 / 0.10 | 11.00 43.00 22.38 9.91 | 100.00 310.00 182.50 81.20 | 16.00 19.00 17.25 1.39 | 1829.24 2690.11 2152.83 260.09 | 980.53 1313.45 1133.75 118.37 | $2.00 \times 10^{22}$ $8.44 \times 10^{22}$ $4.10 \times 10^{22}$ $2.11 \times 10^{22}$ | $2.37 \times 10^{22}$ $1.09 \times 10^{23}$ $4.69 \times 10^{22}$ $2.93 \times 10^{22}$ | $7.49 \times 10^{32}$ $4.40 \times 10^{33}$ $2.07 \times 10^{33}$ $1.25 \times 10^{33}$ | $1.20 \times 10^{33}$ $4.44 \times 10^{33}$ $2.44 \times 10^{33}$ $1.12 \times 10^{33}$ | $4.09 \times 10^{31}$ $5.21 \times 10^{32}$ $3.64 \times 10^{32}$ $1.59 \times 10^{32}$ | $9.31 \times 10^{-3}$ $6.07 \times 10^{-1}$ $2.90 \times 10^{-1}$ $2.33 \times 10^{-1}$ | $9.22 \times 10^{-3}$ $3.78 \times 10^{-1}$ $2.03 \times 10^{-1}$ $1.39 \times 10^{-1}$ |
| Fac | 8 / 4 | 0.15 / 0.10 | 18.00 94.00 38.38 27.32 | 160.00 240.00 205.00 33.81 | 14.00 23.00 17.75 2.60 | 1727.95 2702.99 2201.50 299.48 | 1036.51 1379.29 1249.01 118.72 | $2.02 \times 10^{22}$ $7.25 \times 10^{22}$ $5.08 \times 10^{22}$ $1.71 \times 10^{22}$ | $4.08 \times 10^{22}$ $8.39 \times 10^{22}$ $6.18 \times 10^{22}$ $1.45 \times 10^{22}$ | $6.14 \times 10^{32}$ $4.24 \times 10^{33}$ $2.46 \times 10^{33}$ $1.15 \times 10^{33}$ | $1.10 \times 10^{33}$ $7.23 \times 10^{33}$ $3.28 \times 10^{33}$ $1.84 \times 10^{33}$ | $8.47 \times 10^{31}$ $2.98 \times 10^{33}$ $8.18 \times 10^{32}$ $9.16 \times 10^{32}$ | $2.31 \times 10^{-2}$ $7.87 \times 10^{-1}$ $3.54 \times 10^{-1}$ $2.61 \times 10^{-1}$ | $2.26 \times 10^{-2}$ $4.40 \times 10^{-1}$ $2.39 \times 10^{-1}$ $1.36 \times 10^{-1}$ |
| Fho | 9 / 6 | 0.17 / 0.14 | 5.00 21.00 13.56 5.39 | 270.00 730.00 487.78 163.46 | 16.00 18.00 16.67 0.71 | 2320.46 3261.15 2851.46 301.12 | 1350.17 1733.98 1567.50 119.89 | $2.67 \times 10^{22}$ $4.47 \times 10^{22}$ $3.65 \times 10^{22}$ $5.86 \times 10^{22}$ | $2.69 \times 10^{22}$ $4.86 \times 10^{22}$ $3.68 \times 10^{22}$ $6.81 \times 10^{21}$ | $1.77 \times 10^{33}$ $3.35 \times 10^{33}$ $2.54 \times 10^{33}$ $5.45 \times 10^{32}$ | $1.79 \times 10^{33}$ $3.54 \times 10^{33}$ $2.72 \times 10^{33}$ $5.43 \times 10^{32}$ | $1.55 \times 10^{31}$ $4.17 \times 10^{32}$ $1.79 \times 10^{32}$ $1.42 \times 10^{32}$ | $8.74 \times 10^{-3}$ $2.16 \times 10^{-1}$ $7.55 \times 10^{-2}$ $7.17 \times 10^{-2}$ | $8.66 \times 10^{-3}$ $1.77 \times 10^{-1}$ $6.67 \times 10^{-2}$ $5.91 \times 10^{-2}$ |
| Fhi | 3 / 2 | 0.06 / 0.05 | 15.00 34.00 24.00 9.54 | 380.00 770.00 630.00 217.03 | 16.00 18.00 16.67 1.15 | 3137.14 3287.67 3231.06 81.91 | 1618.36 1869.94 1737.23 126.36 | $3.39 \times 10^{22}$ $4.29 \times 10^{22}$ $3.81 \times 10^{22}$ $4.57 \times 10^{21}$ | $3.95 \times 10^{22}$ $5.26 \times 10^{22}$ $4.82 \times 10^{22}$ $7.54 \times 10^{21}$ | $2.21 \times 10^{33}$ $3.32 \times 10^{33}$ $2.89 \times 10^{33}$ $5.97 \times 10^{32}$ | $2.60 \times 10^{33}$ $3.50 \times 10^{33}$ $3.09 \times 10^{33}$ $4.43 \times 10^{32}$ | $6.67 \times 10^{31}$ $3.92 \times 10^{32}$ $2.00 \times 10^{32}$ $1.70 \times 10^{32}$ | $2.12 \times 10^{-2}$ $1.77 \times 10^{-1}$ $8.05 \times 10^{-2}$ $8.45 \times 10^{-2}$ | $2.08 \times 10^{-2}$ $1.51 \times 10^{-1}$ $7.09 \times 10^{-2}$ $6.98 \times 10^{-2}$ |

| | | | | | | | | | | | | | |
|---|---|---|---|---|---|---|---|---|---|---|---|---|---|
| Fhc | 9 / 6 | 0.17 / 0.14 | 18.00 38.00 29.67 7.47 | 280.00 930.00 561.11 192.97 | 16.00 17.00 16.33 0.50 | 2479.56 3375.79 2890.68 323.62 | 1375.66 1776.88 1551.73 144.33 | $2.89\times10^{22}$ $9.08\times10^{22}$ $4.11\times10^{22}$ $1.92\times10^{22}$ | $3.66\times10^{22}$ $6.73\times10^{22}$ $5.02\times10^{22}$ $8.76\times10^{21}$ | $1.70\times10^{33}$ $4.64\times10^{33}$ $2.46\times10^{33}$ $1.08\times10^{33}$ | $2.22\times10^{33}$ $5.43\times10^{33}$ $2.89\times10^{33}$ $1.09\times10^{33}$ | $3.38\times10^{31}$ $7.94\times10^{32}$ $4.29\times10^{32}$ $2.30\times10^{32}$ | $9.64\times10^{-3}$ $3.34\times10^{-1}$ $2.07\times10^{-1}$ $1.15\times10^{-1}$ | $9.55\times10^{-3}$ $2.51\times10^{-1}$ $1.64\times10^{-1}$ $8.63\times10^{-2}$ |
| Fko | 14 / 10 | 0.26 / 0.24 | 9.00 36.00 15.50 6.39 | 280.00 1080.00 570.71 259.51 | 16.00 28.00 17.64 3.15 | 2344.44 3492.30 2873.83 382.16 | 1321.48 2258.27 1582.40 262.95 | $2.53\times10^{22}$ $9.56\times10^{22}$ $4.53\times10^{22}$ $1.75\times10^{22}$ | $2.17\times10^{22}$ $8.79\times10^{22}$ $5.02\times10^{22}$ $1.89\times10^{22}$ | $1.52\times10^{33}$ $7.53\times10^{33}$ $3.07\times10^{33}$ $1.56\times10^{33}$ | $1.92\times10^{33}$ $7.57\times10^{33}$ $3.40\times10^{33}$ $1.51\times10^{33}$ | $2.76\times10^{31}$ $1.87\times10^{33}$ $3.26\times10^{32}$ $4.71\times10^{32}$ | $5.97\times10^{-3}$ $6.53\times10^{-1}$ $1.40\times10^{-1}$ $1.77\times10^{-1}$ | $5.94\times10^{-3}$ $3.95\times10^{-1}$ $1.06\times10^{-1}$ $1.14\times10^{-1}$ |
| Fki | 24 / 13 | 0.45 / 0.31 | 8.00 44.00 22.21 10.59 | 280.00 1190.00 537.92 235.24 | 16.00 28.00 18.21 3.09 | 2111.07 3533.06 2600.57 390.91 | 1069.61 1936.05 1465.56 260.05 | $2.64\times10^{22}$ $1.13\times10^{23}$ $5.85\times10^{22}$ $2.07\times10^{22}$ | $1.99\times10^{22}$ $9.50\times10^{22}$ $6.17\times10^{22}$ $2.05\times10^{22}$ | $1.41\times10^{33}$ $7.21\times10^{33}$ $3.41\times10^{33}$ $1.45\times10^{33}$ | $1.88\times10^{33}$ $7.26\times10^{33}$ $3.98\times10^{33}$ $1.52\times10^{33}$ | $2.12\times10^{31}$ $3.17\times10^{33}$ $5.70\times10^{32}$ $7.38\times10^{32}$ | $3.33\times10^{-3}$ $1.01\times10^{0}$ $1.99\times10^{-1}$ $2.41\times10^{-1}$ | $3.32\times10^{-3}$ $5.02\times10^{-1}$ $1.42\times10^{-1}$ $1.27\times10^{-1}$ |
| Fkc | 117 / 95 | 2.21 / 2.27 | 11.00 94.00 40.81 17.48 | 250.00 2740.00 929.27 502.81 | 16.00 27.00 18.16 2.30 | 1835.52 3622.89 2796.58 399.27 | 1051.17 3959.81 1836.49 517.70 | $2.10\times10^{22}$ $1.76\times10^{23}$ $7.02\times10^{22}$ $3.23\times10^{22}$ | $4.19\times10^{22}$ $2.17\times10^{23}$ $9.39\times10^{22}$ $3.82\times10^{22}$ | $9.45\times10^{32}$ $2.03\times10^{34}$ $4.75\times10^{33}$ $3.44\times10^{33}$ | $1.28\times10^{33}$ $2.09\times10^{34}$ $5.75\times10^{33}$ $3.72\times10^{33}$ | $2.35\times10^{31}$ $5.02\times10^{33}$ $1.00\times10^{33}$ $8.89\times10^{32}$ | $3.87\times10^{-3}$ $2.04\times10^{0}$ $2.68\times10^{-1}$ $2.67\times10^{-1}$ | $3.85\times10^{-3}$ $6.71\times10^{-1}$ $1.88\times10^{-1}$ $1.20\times10^{-1}$ |